\documentclass{mn2e}
\usepackage{graphicx}

\newcommand{\cm}{\,{\rm cm}}

\newcommand{\cmcube}{\,{\rm cm^{-3}}}

\newcommand{\erg}{\,{\rm erg}}
\newcommand{\FRM}{\,{\rm rad\,m^{-2}}}

\newcommand{\kms}{\,{\rm km\,s^{-1}}}

\newcommand{\kpc}{\,{\rm kpc}}
\newcommand{\Mpc}{\,{\rm Mpc}}
\newcommand{\G}{\,{\rm G}}

\newcommand{\mkG}{\,\mu{\rm G}}

\newcommand{\p}{\,{\rm pc}}
\newcommand{\radm}{\,{\rm rad\,m^{-2}}}

\newcommand{\yr}{\,{\rm yr}}
\newcommand{\Gyr}{\,{\rm Gyr}}
\newcommand{\ini}{{}_\mathrm{i}}

\newcommand{\m}{m}
\newcommand{\M}{M}
\newcommand\mean[1]{\langle{#1}\rangle}
\newcommand{\f}{{}_\mathrm{f}}

\newcommand{\cs}{c_\mathrm{s}}
\newcommand{\cl}{_\mathrm{c}}
\newcommand{\scl}{_\mathrm{sc}}
\newcommand{\Pm}{P_\mathrm{m}}
\newcommand{\Rey}{\mathrm{Re}}
\newcommand{\Reycr}{\mathrm{Re}_\mathrm{cr}}
\newcommand{\Rm}{R_\mathrm{m}}
\newcommand{\Rmcr}{R_\mathrm{m,cr}}

\newcommand{\nel}{n_\mathrm{e}}   
\newcommand{\RM}{\mathrm{RM}}
\newcommand\sfrac[2]{{\textstyle{\frac{#1}{#2}}}}
\newcommand\vect[1]{\bmath #1}

\newcommand{\phzero}{\phantom{0}}
\newcommand{\tphzero}{\phantom{0}\phantom{0}}

\title{Evolving turbulence and magnetic fields in galaxy clusters}
\author[K.~Subramanian, A.~Shukurov and N.~E.~L.~Haugen]%
{Kandaswamy Subramanian,$^{1,4}$\thanks{E-mail: kandu@iucaa.ernet.in (KS);
anvar.shukurov@ncl.ac.uk (AS); nils.haugen@phys.ntnu.no (NELH)}
Anvar Shukurov$^{2,1,4}$ and Nils Erland L.\ Haugen$^{3,5}$\\
$^{1}$Inter-University Centre for Astronomy and
        Astrophysics,  Post Bag 4, Ganeshkhind, Pune 411 007, India\\
$^2$School of Mathematics and Statistics, University of Newcastle, Newcastle
        upon Tyne, NE1 7RU, U.K.\\
$^3$Department of Physics, Norwegian University of Science and Technology,
        H{\o}yskoleringen 5, N-7034 Trondheim, Norway\\
$^4$Isaac Newton Institute for Mathematical Sciences, 20 Clarkson Road,
        Cambridge, CB3~0EH, U.K.\\
$^5$DAMTP, Centre for Mathematical Sciences, Wilberforce Road,
        Cambridge CB3 0WA, U.K.
}

\date{}
\begin{document}

\pagerange{\pageref{firstpage}--\pageref{lastpage}} \pubyear{2005}

\maketitle

\begin{abstract}
We discuss, using simple analytical models and MHD simulations, the
origin and parameters of turbulence and magnetic fields in galaxy clusters.
Any pre-existing tangled magnetic field must decay in a
few hundred million years
by generating gas motions even if the electric conductivity
of the intracluster gas is high.
We argue that turbulent motions can be maintained in the intracluster gas
and its dynamo action
can prevent such a decay and amplify a random
seed magnetic field
by a net factor typically $10^4$ in $5\Gyr$.
Three physically
distinct regimes can be identified in the evolution of turbulence and
magnetic field in galaxy clusters. Firstly,
the fluctuation dynamo will produce microgauss-strong,
random
magnetic fields during the epoch of cluster formation and major mergers. At
this stage pervasive turbulent
flows with r.m.s.\ velocity of about $300\kms$ can be maintained
at scales 100--$200\kpc$. The magnetic field is intermittent,
has a smaller scale of
20--$30\kpc$ and average strength of $2\mkG$.
Secondly, turbulence will decay
after the end of the major merger epoch; we discuss the dynamics of the
decaying turbulence and the behavior of magnetic field in it.
Magnetic field and turbulent speed undergo a power-law decay,
decreasing by a factor of two during this stage, whereas their scales increase
by about the same factor.  Thirdly, smaller-mass
subclusters and cluster galaxies will produce
turbulent wakes where magnetic fields will be generated as well. Although the
wakes plausibly occupy only a small fraction of the cluster volume,
we show that their area covering factor can be close to unity, and thus they
can produce some of the
signatures of turbulence along virtually all lines of sight.
The latter could potentially allow one to reconcile
the possibility of turbulence with ordered filamentary gas structures,
as in the Perseus cluster. The turbulent speeds and magnetic fields in the wakes are
estimated to be of order $300\kms$ and $2\mkG$, respectively, whereas the
turbulent scales are of order $200\kpc$ for wakes behind
subclusters of a mass $3\times10^{13}M_\odot$ and
about $10\kpc$ in the galactic wakes.
Magnetic field in the wakes
is intermittent and
 has the scale of about 30 kpc and 1 kpc in the
subcluster and galactic wakes, respectively.
Random Faraday rotation measure is estimated to be typically 100--$200\radm$,
in agreement with observations.
We predict detectable polarization of synchrotron emission from cluster
radio halos at wavelengths 3--6\,cm, if observed at sufficiently high resolution.
\end{abstract}
\label{firstpage}

\section{Introduction}
Intergalactic gas in clusters of galaxies is
magnetized (see reviews by Kronberg 1994; Carilli \& Taylor 2002; Govoni \& Feretti 2004). The number
of clusters that exhibit detectable synchrotron emission is relatively small,
but it is believed that a magnetic field is present in most clusters (unlike
relativistic electrons). Therefore, a more informative observational tracer
of intracluster magnetic fields
is Faraday rotation of the polarization plane of background radio sources and
central radio galaxies. Clarke, Kronberg \& B\"ohringer (2001) conclude, from
their statistical study of Faraday rotation in 16 galaxy clusters,
that random magnetic fields of a strength (5--10)$\times(l/10\kpc)^{-1/2}\mkG$ (with
$l$ the field scale) permeate the gas within about $500\kpc$ from the cluster
centre; the area covering factor of magnetic fields is close to unity.
The Faraday rotation measures detected are about $200\radm$ for the lines of
sight through clusters' central parts and about $100\radm$ farther out
(see also Clarke 2004; Johnston-Hollitt \& Ekers 2004).
Important constraints on intracluster magnetic fields
come from limits on X-ray emission produced from microwave background
photons by inverse Compton scattering off relativistic electrons
(Bagchi, Pislar \& Lima Neto 1998; Sarazin 1988).
Observational evidence is compatible with a random magnetic field of
r.m.s.\ strength of $1$--$10\mkG$ and coherence length of about
10--$20\kpc$.

Faraday rotation maps of a number of
radio galaxies in clusters have also been
analyzed. Eilek and Owen (2002)
studied Faraday rotation maps
of radio sources in the centres of the Abell clusters A400 and A2634,
and found patches of RM fluctuations
on scales
about 10--$20 \kpc$.
Assuming this to be also
the
coherence
scale for the field, these authors deduce
field strengths of
1--$4 \mkG$.
Vogt and En{\ss}lin (2003, 2005), using a novel technique
(En{\ss}lin \& Vogt 2003),
estimate magnetic field strength to be $3 \mkG$ in A2634, $6 \mkG$
in A400 and $7 \mkG$ in Hydra~A. They obtain field correlation
length of $4.9 \kpc$, $3.6 \kpc$ and $3 \kpc$ for these clusters,
respectively.
It cannot be excluded, however, that the Faraday rotation of cluster
radio sources is contaminated by that arising in dense turbulent
cocoons around the radio galaxies, rather than in the
intracluster medium proper (Rudnick and Blundell 2003;
see En{\ss}lin et al.\ 2003 for another view).
For this reason, the statistical studies of Faraday
rotation of background radio sources, referred to above,
provide perhaps a more convincing evidence for
cluster-wide magnetic fields and their properties.

The origin of the cluster magnetic fields remains unclear.
Carilli \& Taylor (2002) (see also Tribble 1993a)
argue that the small value of electric resistivity of the intracluster plasma
guarantees that the decay time of magnetic field will be comparable to or
exceed the cluster lifetime. They conclude that any magnetic
field (e.g., that captured by a
cluster during its formation) would survive for a long time.
However, any inhomogeneous magnetic field will drive
motions via the Lorentz force, and the motions will decay, plausibly in the form
of decaying MHD turbulence (e.g., Biskamp 2003; see below). The turbulent decay time is comparable
to the eddy turnover time of the largest eddies,
about
$10^8\yr$, irrespective of the resistivity or viscosity of the gas.
(Here we have taken a coherence scale of order $10$ kpc for the
field and the induced motions, with associated
turbulent velocities of order $100\kms$).
Although the energy
density in MHD turbulence decays
with time
 as a power law,
this time scale is still much
shorter
 than the
typical age of a cluster, which is thought to be
several billion years.
Therefore, one has to provide explicit explanation of the
origin and persistence of magnetic
fields in the clusters; reference to the low Ohmic resistivity of the
intracluster plasma is not sufficient if the gas is turbulent or the
magnetic field is tangled.

An obvious option to explain intergalactic magnetic fields is to consider
magnetic fields stripped from galaxies. Since the intracluster gas is enriched
with metals, at least part of it originates in galaxies (Sarazin 1988).
However, the strength of magnetic field produced by the stripping cannot exceed
$\simeq0.1\mkG$ even in the cores of rich clusters and even if spiral galaxies
with relatively strong large-scale field are involved (see Appendix~\ref{seed}).
Another possibility is that the intracluster field is supplied by active
galaxies within the cluster. As we discuss in Appendix~\ref{seed}, this
mechanisms can provide relatively strong magnetic field but fails to explain
how the field can be maintained against turbulent decay. In addition, it is not quite
clear how efficiently the magnetized relativistic plasma of the radio lobes
can be mixed with the thermal intergalactic plasma and what would be the resulting
scale of magnetic field.
Altogether, the above mechanisms can only provide suitable seed magnetic field for
the dynamo action in the intracluster plasma.

In most astrophysical systems, like disc galaxies, stars and planets, rotation
is crucial for maintaining their magnetic fields, both by providing strong
shear and by making (when coupled with stratification) random flows helical,
and hence leading to mean-field dynamo action. However, galaxy clusters are
believed to have fairly weak rotation (if any at all), so one has to appeal
to some other mechanism for understanding cluster magnetism.

Another possibility to generate magnetic fields is related to the
fluctuation dynamo action (Batchelor 1950; Kazantsev 1967),
where random flow of electrically conducting fluid
generates random magnetic field (Zeldovich, Ruzmaikin \& Sokoloff 1990).
This mechanism does not require any
rotation or density stratification, and only relies on the random nature
of the flow;  so fluctuation dynamos can be active in virtually any turbulent
environment where the plasma is ionized.
Apart from the randomness of the flow, it is required that the magnetic Reynolds
number is large enough, i.e., that the electric conductivity is high enough,
and/or plasma motions are sufficiently intense, and/or their scale is sufficiently
large.

The earliest theories of intracluster
magnetic fields were based, implicitly or explicitly, on
fluctuation dynamo theory under
the assumption that galaxy clusters are steady-state turbulent
systems (Jaffe 1980; Roland 1981; Ruzmaikin, Sokoloff
\& Shukurov 1989; Goldman \& Rephaeli 1991; De Young 1992).
The source of turbulence adopted by several authors were
turbulent wakes of the cluster galaxies. This picture has to be reconsidered
for several reasons.

Firstly, turbulence from galactic wakes can fill the cluster volume only
if the effective galactic radius is of order $10\kpc$
(e.g., Ruzmaikin et al.\ 1989), i.e.,
if the interstellar gas is not stripped by the ram pressure
of the intracluster gas. If the gas stripping is complete, the wake is only
produced by gravitational accretion (Bondi 1952), and its radius is about the
accretion radius $r_\mathrm{g}={2GM}/({\cs^2+V^2})$
where $\cs$ is the speed of sound, $V$ is the galactic speed,
$M$ is the galactic mass, and $G$ is Newton's gravitational
constant. For $\cs=10^3\kms$, $V\approx\cs$ and
$M=10^{11}M_\odot$, the gravitational accretion radius $r_\mathrm{g}\simeq0.5\kpc$ is much smaller
than both the galactic radius and the apparent scale of the random magnetic
field in the intergalactic gas (both usually assumed to be of order $10\kpc$).
Hence, sufficiently strong, volume-filling turbulent wakes whose width is
comparable to the galactic size can only arise if
the galaxies retain significant amounts of their interstellar gas. If the
stripping of interstellar gas by ram pressure is efficient, galactic wakes
are rather weak (Portnoy, Pistinner \& Shaviv 1993; Balsara, Livio \& O'Dea
1994;  Acreman et al.\ 2003 and references therein; see however Toniazzo \&
Schindler 2001 who argue that the stripping efficiency is exaggerated in the
above papers). Therefore turbulence generated in galactic wakes may not fill
the cluster volume (see also Sect.~\ref{GW}).

Further, numerical simulations of De Young (1992), using a closure
model, gave pessimistic
estimates for magnetic fields produced by the dynamo when the turbulence is induced by galactic
wakes. However, it is not clear if the resolution of those simulations (i.e., the
effective magnetic Reynolds number) was high enough to obtain dynamo action,
so this objection to the dynamo models is questionable. Recent direct simulations
of dynamo action in turbulent flows (Haugen, Brandenburg \& Dobler
2003, 2004; Schekochihin et al.\ 2004) have confirmed the efficiency of dynamo
action in random non-helical flows.

Perhaps more importantly, cluster dynamics has been reconsidered
recently, and the emerging picture is very different in that clusters
may not be relaxed systems, but still remain in the state of formation via
major mergers and accretion of smaller-mass subclusters.
Numerical simulations and recent observations
strongly suggest that a random flow, perhaps of turbulent
nature, can be maintained for a few crossing times of the forming galaxy
cluster (Norman \& Bryan 1999). Roettiger et al.\ (1999a,b) have
found that magnetic field can be amplified by these motions
(see also results obtained with SPH simulations by Dolag, Bartelmann \& Lesch
1999, 2002). However, the resolution of
the simulations is still poor, and quantitative estimates of the turbulence
parameters and especially of its effects on magnetic field are very uncertain.

In addition to the volume-filling flow produced during the cluster formation,
significant random flows can still be generated in wakes behind
infalling subclusters and cluster galaxies. These
are not expected to fill the cluster volume, but we argue below that they can
have significant area covering factor.
There is also a possibility that radio galaxies can stir the
intracluster gas as their plasma buoyantly rises through the gas
(Br\"uggen et al.\ 2002; En{\ss}lin \& Heinz 2002).
Another possible consequence of  radio galaxy jets and/or lobes propagating
at subrelativistic speeds through the cluster plasma, is
the generation of turbulence in a cocoon surrounding the radio source
(see for example Reynolds, Heinz \& Begelman 2002).

The content of the paper is as follows.
We consider the evolution of turbulence in the intracluster gas of
a galaxy cluster during and after its formation in Sect.~\ref{CFIT}.
Random flows produced during the merger epoch can lead to magnetic
field generation via the fluctuation dynamo  as discussed in Sect.~\ref{MFDT}.
In  Sects~\ref{DT} and \ref{SFD} we present evidence that the
decay of both magnetic and kinetic energies after the epoch of
major mergers will be a power law in
time, rather than exponential, because
of the turbulent nature of the flow.
During the decay phase of the turbulence, the correlation scale of the magnetic
field will grow (Frisch 1995; Olesen 1997; Biskamp \& M\"uller 1999;
Christensson, Hindmarsh \& Brandenburg 2001), which slows down the decay of
the Faraday rotation measure produced in the
intracluster gas (Sect.~\ref{ODFRM}).
In Sect.~\ref{MMTW} we argue that significant
turbulence and magnetic field amplification
can occur in turbulent wakes of smaller-mass
subclusters and cluster galaxies. The
turbulent wakes may not fill the volume, but can cover the cluster's projected
area (Sect.~\ref{CFTW}). We present order-of-magnitude
estimates of the parameters of the random flows at various stages of the
cluster evolution in Sect.~\ref{CFIT}, and of magnetic fields generated by the
flows, in Sect.~\ref{MFDT}, which are further substantiated
by numerical simulations
of dynamo action in driven and decaying random flows
discussed in Sect.~\ref{SFD}.
Faraday rotation measure and polarized radio emission produced by these
magnetic fields are discussed in Sect.~\ref{OD}. Our results are
summarized in Sect.~\ref{Disc} and Table~\ref{res}.

\section{Intracluster turbulence}\label{CFIT}

An upper limit on the turbulent velocity in a steady-state galaxy cluster
follows from the
requirement that the rate of dissipation of the turbulent energy should not
exceed the X-ray luminosity of the cluster $L_X$, i.e., $\frac12 v_0^3/l_0\la
L_X/M_\mathrm{g}$, where $v_0$ and $l_0$ are the turbulent speed and scale,
respectively, and $M_\mathrm{g}$ is the mass of the intergalactic gas. This
yields
\begin{equation}        \label{upperlimit}
v_0\la 180\,\frac{\mathrm{km}}{\mathrm{s}}
\left(\frac{l_0}{200\kpc}\right)^\frac{1}{3}\!\!
\left(\frac{L_X}{10^{45}\erg/\mathrm{s}}\right)^\frac{1}{3}\!\!
\left(\frac{M_\mathrm{g}}{10^{14}M_\odot}\right)^\frac{1}{3}\!\!.
\end{equation}
This restriction applies to a steady state, and stronger turbulence can be
driven in an evolving cluster, where the energy released by the decay of
turbulent motions heats up the gas. Therefore, turbulent velocities
significantly exceeding the above value can be considered as an
indication of the cluster's ongoing evolution. The turbulent nature of the flow
is important, however: Eq.~(\ref{upperlimit}) only applies to turbulent flows
and has to be reconsidered in the case of a random flow without turbulent energy cascade.
As we argue below, turbulent motions during the epoch of
cluster formation  are intense enough to violate the constraint (\ref{upperlimit})
because they keep
evolving and their dissipation contributes to the heating of the intracluster
gas to the virial temperature.

We now discuss various sources of turbulence in galaxy clusters.

\subsection{Turbulence produced during cluster formation}\label{TPDCF}

Theories of hierarchical structure formation suggest that clusters of galaxies
have been assembled relatively recently. $N$-body simulations indicate that
the clusters form at the intersection of dark matter filaments in the large-scale
structure, and result from
both major mergers of objects of comparable mass (of order
$10^{15}M_\odot$) and  the accretion of smaller clumps onto massive
protoclusters. It is likely that intense random vortical flows, if not
turbulence, are produced in the merger events (Kulsrud et al.\ 1997; Norman \&
Bryan 1999; Ricker \& Sarazin 2001). These would originate not only due to
vorticity generation in oblique accretion shocks and instabilities during the
cluster formation, but also in the wakes of the smaller clumps.
In this
section we summarize some of the work on the cluster-wide turbulence resulting
from major mergers. Its consequences for the generation of cluster magnetic
fields are considered in Sect.~\ref{MFDT}.

There have been several numerical simulations of gas dynamics during the
formation of galaxy clusters. Norman \& Bryan (1999) find that the
intracluster medium becomes turbulent during cluster formation, with turbulent
velocities of about $400\kms$ within $1\Mpc$ from the centre of a cluster and
eddy sizes ranging from 50 to 500\kpc; the random flow is volume
filling (see also Sunyaev, Norman \& Bryan 2003). In the cluster merger model
of Ricker \& Sarazin (2001), ram pressure displaces gas in the cluster core
from the bottom of the potential well. The resulting
convective plumes produce large-scale disordered motions with eddy size up to
several hundred kiloparsecs; even after 15\,Gyr of evolution,
turbulent velocities in the inner parts
remain at a surprisingly high level of 10--40\% of the sound speed (i.e.,
100--$400\kms$), and grow up to the sound speed in the outer parts.

Observational evidence of intracluster turbulence is scarce. From analysis of
pressure fluctuations as revealed in X-ray observations, Schueker et al.\
(2004) argue that the integral turbulent scale in the Coma cluster is close to
100\,kpc, and they assume a turbulent speed of $250\kms$ at that scale.  The
nonthermal broadening, by the turbulence, of X-ray spectral lines of ionized
iron can be detectable with future X-ray observatories (Inogamov \& Sunyaev
2003).

The spatial resolution of all available simulations is rather coarse (of order
$10\kpc$ or worse) and only comparable with the apparent scale of the
intracluster magnetic field. Even if the random nature of the resulting flow
is obvious, it is not clear if it will evolve into developed turbulence, i.e.,
if the turbulent cascade is established to carry energy to small scales where it
dissipates into heat. From the viewpoint of magnetic field generation, the
presence of turbulence as such is not required; the randomness of the flow is
sufficient (Kazantsev 1967; Zeldovich et al.\ 1990).
However, the dynamics of the flow and its evolution do depend on
whether or not the turbulent cascade persists (see Sect.~\ref{DT}).

The flow
can become turbulent if the kinematic Reynolds number  in the intracluster
gas, $\Rey$, is large enough. Following Sarazin (1988) and Ricker \& Sarazin
(2001), an estimate of $\Rey$ can be obtained as
\begin{equation}\label{Reyn}
\Rey=\frac{v_0 l_0}{\nu}
        \simeq 3\frac{v_0l_0}{\cs\lambda\delta}
        =3{\cal M}\frac{l_0}{\lambda\delta}\;,
\end{equation}
where $\nu=\frac13\cs\lambda\delta$ is the effective kinematic viscosity, $\cs$ is the speed
of sound (assumed to be close to the thermal speed),
${\cal M}$ is the Mach number, $\lambda$ is the ion mean free
path, and subscript `0' refers to the energy-range values.
Here we have introduced parameter $\delta$ that quantifies the poorly understood
behaviour of viscosity in the intracluster plasma. The standard, Spitzer's value
of $\lambda$ can be written as
\[
\lambda\simeq5\kpc\left(\frac{\cs}{10^3\kms}\right)^4
        \left(\frac{\nel}{10^{-3}\cm^{-3}}\right)^{-1},
\]
with $\nel$ the electron number density.  For
${\cal M}\simeq1$, $l_0\simeq100$--$500\kpc$ and $\lambda\ga1\kpc$, this
yields $\Rey\la(300$--$1500)\delta^{-1}$. This estimate is, however, suspect because the mean
free path is comparable to the scale of inhomogeneities in the gas. It is also
clear that even a weak
seed intracluster magnetic field could strongly reduce the effective
viscosity and make it anisotropic. The effective Reynolds number can be
significantly larger if any shorter length scale plays the role rather than
the Coulomb mean free path, or a frequency higher than the ion collision
frequency (these may be associated with plasma instabilities and/or waves).
Schekochihin et al.\ (2005b) argue that the firehose and/or mirror
instabilities can provide the effective diffusion in the magnetized plasma;
the corresponding length scale is the ion gyroradius, a quantity normally much smaller than
the mean free path. These uncertainties
are allowed for, in a heuristic manner,  by choosing $\delta<1$;
Fabian et al.\ (2005) scale their results by $\delta=0.1$.
This prescription can be an oversimplification as it omits plausibly
important physical effects arising from the anisotropy of viscosity in
magnetized plasma (cf.\ Schekochihin et al.\ 2005b). However,
even if one cannot readily provide a confident estimate of $\Rey$,
Eq.~(\ref{Reyn}) suggests
that it will be large enough as to ensure that random motions driven by major
merger events can become turbulent. Only this fact is important for our purposes in this paper.
Our simulations of a flow driven by
external random force (Sect.~\ref{SFD}) have been performed for
$\Rey=100$--400, and they do indeed show a flow with broad range of scales
typical of turbulent flows (and so inertia forces dominate over viscosity),
even if the inertial range is not wide at these modest Reynolds numbers.

To summarize, the above results seem to converge to the following picture.
Random motions driven in major merger events have the typical initial
speed of $v_0\ini\simeq300\kms$ and scale $l_0\ini\simeq100$--$200\kpc$,
so that the turnover time of the energy-range eddy is
$t_0\ini=0.3$--$0.6\Gyr$. The random motions will be maintained at this level
during the major merger epoch, whose duration can be as large as
$t\f\simeq3$--$5\Gyr$.  (The notation is motivated in Sect.~\ref{DT}.)

\subsection{Dynamics of decaying turbulence}\label{DT}

The random flows produced by major mergers will not remain statistically
steady after the end of the merger event.
Unlike a laminar flow that decays exponentially in time due to
viscosity, turbulent kinetic energy decays slower, as a power law (e.g.,
Landau \& Lifshitz 1975, Frisch 1995). The reason for this is that kinetic
energy mainly decays at small scales, to where it is constantly supplied by the
turbulent cascade. As a result, the energy decay rate depends nonlinearly on
the energy itself, which makes the decay a power law in time. (The
corresponding calculation is provided below.)
Our simulations of Sect.~\ref{SFD} confirm that the power-law decay occurs
even for the Reynolds number as small as $\Rey\approx100$.

\begin{figure}
\centerline{\includegraphics[width=0.35\textwidth]{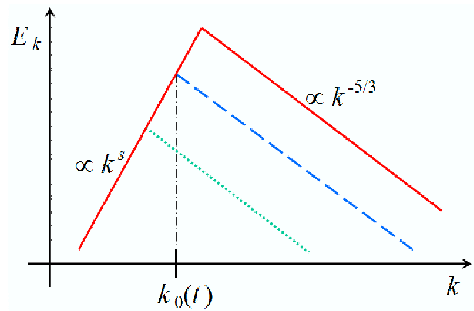}}
\caption{\label{spectrum_decay}
The evolution of the spectrum of decaying turbulence, where the
spectral exponent remain constant both in the inertial range, where it is
equal to $-5/3$, and at large scales  where it is equal to $s$. Solid, dashed
and dotted lines show the spectrum at
consecutive times.  As the energy dissipates, the energy-range wave number
$k_0$ reduces.}
\end{figure}

In this section we review simple models of decaying
hydrodynamic turbulence. The effects of magnetic field on the flow can be
neglected at early stages when magnetic field is still weak. In Sect.~\ref{SFD}
we present numerical simulations where the effects of magnetic field on the
flow are fully allowed for.

Consider an initial spectrum of turbulence shown with solid line in
Fig.~\ref{spectrum_decay}, where $E_k = C k^s$ (with $s > -1$) at scales
$k < k_0$, and $E_k \propto k^{-5/3}$ at smaller scales as in Kolmogorov
turbulence. Here $kE_k$ is the specific energy  per unit
logarithmic interval in the $k$-space. It is related to
the turbulent velocity at wave number $k$ via
$v_k \propto (kE_k)^{1/2}$;
$E_k$ has a maximum at a certain wave number
$k_0$, which is therefore called the energy-range wave number.
Turbulent flow remains in a statistically steady state despite viscous
dissipation at small scales if it is driven at larger scales.
When such a driving ceases, the turbulence decays.

Motions at small scales are the first to be affected by viscosity.
It is reasonable to expect
that the exponent $s$ and constant $C$ are preserved during the decay (e.g.,
Sect.~7.7 of Frisch 1995). Consider the time-dependent total specific
turbulent energy
$E(t)$, which is approximately equal to that at the energy-range scale,
$E(t)\simeq\frac{1}{2}v_0^2$, if the inertial part of the spectrum
is steep enough. On
the other hand, $E(t)=\int_0^\infty E_k\,dk \propto k_0^{s+1}$.
As long as the Reynolds number remains large,
and so viscosity at $k_0$ is negligible, energy at the energy-range
wave number $k_0(t)$ mainly decays because it cascades
to smaller scales. Hence,
the energy loss rate is given by $dE/dt=-v_0^2/t_0\simeq -v_0^3k_0$, where
$t_0\simeq (v_0k_0)^{-1}$. Since
$v_0\propto E^{1/2}$ and $k_0\propto E^{1/(s+1)}$, the evolution of
the total turbulent energy is governed by
\begin{equation}
\frac{d E}{dt} \simeq -A E^{(3s+5)/[2(s+1)]}\;,
\label{dec}
\end{equation}
where $A$ is a certain constant related to $C$. This equation can easily be
integrated (for constant $C$) to yield (see Fig.~\ref{spectrum_decay})
asymptotic forms (applicable at large $t$) for the energy decay law
\[
E(t) \propto t^{-\alpha}\;,
\]
and for the energy-range scale growth law
\[
l_0 \propto k_0^{-1}\propto t^\beta\;,
\]
where
\[
\alpha = 2\frac{s+1}{s+3}\;,
\qquad
\beta = \frac{2}{s+3}\;.
\]
The value $s=2$ gives a `white noise' spectrum at large scales, where the
three-dimensional spectrum $k^{-2}E_k$ is flat. In this case
\[
\alpha = 6/5\;,
\qquad
\beta = 2/5\;.
\]
Another possibility often considered in this context is $s=4$, where one gets
$\alpha = 10/7$ and $\beta = 2/7$ (Skrbek \& Stalp 2000; Touil, Bertoglio \&
Shao 2002).

There are arguments suggesting that $s=2$ is a better acceptable value than
$s=4$; in particular, the coefficient $C$ is time-dependent
for $s=4$, and this makes the energy decay significantly slower than that
obtained above (Frisch 1995).
Also $s=4$ would be relevant for an incompressible flow, whereas
the cluster flows are expected to be mildly compressible, and $s=2$
could be more appropriate.
The turbulence decay is sensitive to the
detailed physical nature of the system, and it is often slower
than derived above.
The decay of the MHD turbulence can be significantly slowed
down if the system has
non-zero invariants such as magnetic helicity and/or cross-helicity (Biskamp
2003). If the intracluster seed magnetic fields are due to stripping of
the galactic magnetic fields, then they may have both types of helicities.
The decay law is also sensitive to the relation
between the turbulent energy-range scale and the size of the system; the decay
speeds up to become $E\propto t^{-2}$ when the two scales become
comparable and the value of $k_0$ cannot decrease any further (Skrbek \& Stalp
2000; Touil et al.\ 2002).

We adopt $\alpha=6/5$ and $\beta=2/5$ for numerical estimates in what follows,
so that
\begin{equation}\label{Ekdecay}
E\simeq \sfrac{1}{2}v_0^2\propto \left(\frac{t-t\f}{t_0\ini}\right)^{-6/5},
\quad
k_0\propto \left(\frac{t-t\f}{t_0\ini}\right)^{-2/5},
\end{equation}
for $t-t\f\gg t_0\ini\,,$
where subscript `i' refers to the start of the evolution, $t_0\ini$ is a
certain dynamical time scale, which can be identified with the initial
turnover time of the energy-containing eddies, $t_0\ini=l_0\ini/v_0\ini$,
subscript `0' refers to the
energy-range (correlation) scale of the motion (which varies with time),
and the decay starts at time $t=t\f$ when the flow forcing ceases.
This decay is faster than in many other models of decaying turbulence;
thus, our conclusions will be rather conservative with respect to the
intensity of turbulence at late times.

With the above decay law of turbulence, the Reynolds number evolves slowly as
\[
\Rey\propto \left(\frac{t-t\f}{t_0\ini}\right)^{-1/5}\;,
\qquad t-t\f\gg t_0\ini\;.
\]
Allowing for the initial period
$t\f=3\Gyr$ of sustained
turbulence, $\Rey$ decreases only by a factor of 1.4 after the total
evolution time of $t=6\Gyr$ for
$l_0\ini=150\kpc$ and $v_0\ini=300\kms$ (yielding $t_0\ini=0.5\Gyr$).

\subsection{Minor mergers and turbulent wakes}\label{MMTW}

Consider the infall of relatively small subclusters of mass $\m$ into an
already formed cluster of mass $\M$.
Define
$d^2p/(d(\ln m)\,dt)$
as the probability that in
a time $dt$ a subcluster,
whose mass belongs to
a logarithmic mass interval
$[\ln m, \ln m +d(\ln m)]$, merges with the bigger cluster of
mass $M$.
The merger rate $d^2p/(d(\ln m)dt)$ scales with the
subcluster mass as (Lacey \& Cole 1993)
\begin{equation}\label{mergerrate}
\frac{d^2p}{d(\ln m)\, dt}\propto m^{-1/2}\quad\mbox{for } \frac{m}{M}\ll1\;.
\end{equation}
Thus, the merger rate of masses of order $10^{13}M_\odot$ is
about 10 times larger than that for $10^{15}M_\odot$. If major mergers of
masses of order $10^{15}M_\odot$ occur once in $3\Gyr$, the time interval
between mergers with $10^{13}M_\odot$ subclusters will then be of the order of
$0.3\Gyr$ (see also Norman \& Bryan 1999).
Such minor mergers are thought to play an
important role in explaining the observed cold fronts
in clusters (Heinz et al.\ 2003; Motl et al.\ 2004 and references therein).
They can also generate turbulence in the wake of a
moving subcluster (cf.\ recent simulations of Takizawa 2005).
Turbulence generated by subclusters was suggested
by Norman and Bryan (1999) to be a major source of the random motions
observed in their simulations of cluster formation.
Here we examine this issue further with analytical estimates, using
parameters of galaxy clusters and smaller structures obtained from
hierarchical theories of structure formation.

\subsubsection{Ram pressure stripping}

The subclusters contain gas which can be partially stripped by hydrodynamic
interaction with the cluster gas (by ram pressure stripping and via
hydrodynamic instabilities) (Fabian \& Daines 1991;
Acreman et al.\ 2003). A simple
criterion for the radius $R_0$ within which the subcluster gas remains
unstripped can be obtained as follows. The
ram pressure force exerted on a gas sphere of radius $R_0$ is
equal to $\rho\cl v\scl^2\pi R_0^2$, where $\rho\cl$ is the intracluster gas
density and $v\scl$ is the speed at which the subcluster moves through the
intracluster gas.
Following Fabian \& Daines (1991), we note that the
gravitational restoring force per unit area due to the subcluster mass
is comparable to the gas pressure in the subcluster, assuming that the
subcluster is in hydrostatic equilibrium. The gas sphere will be removed from
the subcluster if the restoring force is smaller than the ram pressure force.
Thus, a local criterion for retaining the gas at a distance $R_0$
from the subcluster centre is
\begin{equation}        \label{strip_ineq}
\rho\cl v\scl^2 \leq f  \rho\scl(R_0) u^2\;,
\end{equation}
where $\rho\scl(R)$ is the gas density distribution of the
subcluster, $R$ is the subcluster's spherical radius, $f$ is a numerical
factor of order unity, and $u$ is the gas velocity dispersion within the
subcluster. We adopt, for illustrative purposes, gas density profiles for the
cluster and subcluster, respectively, of the form
\[
\rho\cl(r) =  \frac{{\rho\cl}_0}{ [1 + (r/r\cl)^2]}\;,
\qquad
\rho\scl(R) =  \frac{{\rho\scl}_0}{ [1 + (R/R\scl)^2]}\;,
\]
where ${\rho\cl}_0$ and ${\rho\scl}_0$ are the respective central gas densities
and $r\cl$ and $R\scl$ are the corresponding gas core
radii. (These correspond to the standard $\beta$-profile with
the slope parameter $\beta=2/3$ -- Sarazin 1988).
From Eq.~(\ref{strip_ineq}),
the subcluster gas is retained at radii smaller that $R_0$, where
\begin{equation}
\left(\frac{R_0}{R\scl}\right)^2
= f\frac{{\rho\scl}_0 u^2}{{\rho\cl}_0 v\scl^2}
\left[1 + \left(\frac{r}{r\cl}\right)^2\right] -1 \;.
\label{strip}
\end{equation}
Takeda, Nulsen \& Fabian (1984) suggest $f \simeq 2$.

Parameters of clusters and subclusters that enter Eq.~(\ref{strip}) vary
broadly in both observed and simulated clusters. Suitable values can be
selected as follows. For example, consider subclusters predicted by the
hierarchical theory of structure formation (Peebles 1980;
Padmanabhan \& Subramanian 1992; Padmanabhan 1993).
Suppose that the initial density fluctuations can be described as a Gaussian random
field with the r.m.s.\ density contrast $\sigma_m(m)$, where
$m$ is the mass of the structure. In the hierarchical theory,
$\sigma_m(m) \propto m^{-(3+n)/3}$ with $n$ close to $-1$ at the cluster
scales and to $-2$ at the galactic scales. For a density
fluctuation which is $\mu$ times the above r.m.s.\ value,
the following scaling laws can be obtained:
$r_\mathrm{vir}\propto \mu^{-1} m^{(n+5)/6}$ for the virial radius;
$u^2\simeq Gm/r_\mathrm{vir}\simeq \mu m^{(1-n)/6}$ for the virial velocity;
and $\rho\propto m/r_\mathrm{vir}^3\propto \mu^3 m^{-(n+3)/2}$
for the average gas density. This suggests the average pressure scaling
\[
\rho\scl u^2 \propto \mu^4 m^{-2(n+2)/3}\;.
\]

We adopt $n = -1.5$, $M=10^{15}M_\odot$,
$m = 3 \times 10^{13} M_\odot$, and a bulk velocity of the subcluster
of order the cluster velocity dispersion, $v\scl\simeq1000\kms$.
For comparison, the merging components of the Coma cluster have virial masses
$0.9\times10^{15}M_\odot$ and $6\times10^{13}M_\odot$ (Colless \& Dunn 1996).
We also assume that the cluster and subcluster correspond to
density fluctuations of the same value of $\mu$.
Then Eq.~(\ref{strip}) yields $R_0=2.3 R\scl$ at the
cluster centre, $r=0$, and
$R_0= 3.4 R\scl$ at the cluster core
radius, $r = r\cl$. So, according to this criterion, gas within 2--3
subcluster core radii will not be stripped as the subcluster falls, along a
radial orbit, into a cluster which is about $30$ times larger in mass.

Further, we take the gas core radius to be proportional to the virial radius.
Indeed, Sanderson and Ponman (2003) suggest that the gas core radius is
about $0.1 r_\mathrm{vir}$ for clusters with temperature
exceeding 1\,keV, or the mass of a few times $10^{13} M_\odot$.
Then the subcluster
gas core radius is about $(m/M)^{(n+5)/6} \approx 0.13$
times the cluster gas
core radius for $m/M=0.03$.
For a rich cluster with the virial radius $3\Mpc$ and the
core radius ten times smaller, or $r\cl = 300\kpc$,
we obtain the subcluster gas core radius as $R\scl \simeq 40\kpc$.
This implies for such subclusters the stripping radius of at least
\[
R_0 \simeq 100\kpc\;.
\]
We adopt these values for
qualitative estimates, keeping in mind that scatter about the fiducial values
is likely to be large.

Heinz et al.\ (2003) simulated a subcluster with a shallower
gas density profile ($\beta=0.5$), a somewhat large core radius
$R\scl= 250\kpc$, central gas number density
${\rho\scl}_0 = 3.6 \times 10^{-3}\cm^{-3}$ and
a temperature $T = 3.2$\,keV, moving through a uniform gas
of a density ${\rho\cl}_0 = 4.6 \times 10^{-4}\cm^{-3}$ and temperature
$T = 7.7$\,keV. These authors find that the subcluster gas within
$R_0 \simeq 2 R\scl$ survives ram pressure stripping, which compares favourably
with our estimates based on Eq.~(\ref{strip_ineq}).

Flow past a solid sphere develops
into a turbulent wake for sufficiently large
Reynolds numbers. Experiments and numerical
simulations (Tomboulides \& Orszag 2000 and references therein) show that
the transition to turbulence occurs at
$\Rey \approx 400$, via the Kelvin--Helmholtz
instability of a shear layer that results from the separation
of the boundary layer on the sphere's surface. It is not clear what is the critical
Reynolds number for a gaseous sphere. It can be speculated that the
entrainment of the dense subcluster gas into the flow can be a cause
of the flow randomness
additional to that past a solid sphere.
The Kelvin--Helmholtz instability does indeed develop on
the boundary between the subcluster and the ambient gas, e.g., in
the simulations of Takizawa (2005) among many other authors, leading to prominent
eddy-like structures in the subcluster wake that can be described as a turbulent flow.

Nulsen (1982) describes how the introduction of eddies of a scale $l$ can make
the boundary layer smooth on this scale, suppressing the Kelvin--Helmholtz
instability at wavelengths smaller than $l$. Longer-wavelength modes are still
unstable, and the largest unstable scales are comparable to the stripping
radius. The Kelvin--Helmholtz instability is efficient in eventually removing
gas from the subcluster. According to Heinz et al.\ (2003), all the gas is
removed after a time of order a few times $10 R_0/v_c$, that is a few billion
years. This implies that a subcluster can generate a turbulent wake during one
or two passages through the cluster.

Altogether, the flow produced by the Kelvin--Helmholtz instability can produce
turbulence in the subcluster's wake, subject to the same reservations as
discussed after Eq.~(\ref{Reyn}).  It is then plausible that the wake far
downstream of the subcluster is well described by Prandtl's self-similar
solution for turbulent wakes.

\subsubsection{The area covering and volume filling factors of turbulent wakes}\label{CFTW}

Prandtl's solution for the turbulent scale and velocity variation with distance $x$
along the wake has the form (Landau \& Lifshitz 1975)
\begin{equation}                \label{lv}
l_{0x}\simeq L\ini(x/L\ini)^{1/3}\;,\quad v_{0x}\simeq V\ini(x/L\ini)^{-2/3}\;,
\end{equation}
where the turbulent velocity near the head of the wake can be identified with
some fraction of the subcluster speed, $V\ini\simeq v\cl\simeq1000\kms$; and the
initial value of the turbulent scale $L\ini$ is close to the stripping
radius $R_0$ estimated above, $L\ini\simeq R_0$.
The Reynolds number varies along the wake as
\[
\Rey(x)=\Rey\ini(x/R_0)^{-1/3}\;,
\qquad
\Rey\ini=\frac{L\ini V\ini}{\nu}\;,
\]
and we assume that the wake remains turbulent as long as the Reynolds number
exceeds a certain critical value, $\Reycr$.
Then the length of a turbulent wake, $X$, follows from $\Rey(X)=\Reycr$ as
\begin{equation}\label{XR0a}
\frac{X}{R_0}\simeq \left(\frac{\Rey\ini}{\Reycr}\right)^3\;,
\end{equation}
where $\Reycr=400$ (Tomboulides \& Orszag 2000) can be adopted for illustrative purposes.
Thus, we assume that the critical value of $\Rey$ required to maintain turbulence
within the wake is the same as that to produce it immediately behind a solid
sphere. This will result in a
quite conservative estimate of the wake length since turbulence can plausibly
sustain at even smaller local values of $\Rey$.

The area of a single wake of a length $X$ seen from the side is given by
\[
S=2\int_{ R_0}^{X} R_0(x/R_0)^{1/3}\,dx=\frac32 R_0^2
\left[\left(\frac{X}{R_0}\right)^{4/3} - 1\right].
\]
If the wake axis is inclined by a random angle $\alpha$ to the line of sight, where
$\alpha$ is uniformly distributed with $0\leq\alpha\leq\pi$,
the average area of a single wake in the sky plane is given by
\[
\overline{S}=\xi S\;,\qquad
\xi=\frac{1}{2}\int_0^\pi\sin^{7/3}\alpha\,d\alpha\
\approx0.74\;.
\]
The area covering factor of $N$ wakes within a region larger in diameter $2r$ than
the wake length, $r>X/2$,  follows as
\begin{equation}\label{fS}
f_S=\frac{N\overline{S}}{\pi r^2}
\simeq  \frac{3}{2\pi}N \xi
        \left(\frac{R_0}{r}\right)^2
        \left[\left(\frac{X}{R_0}\right)^{4/3} - 1\right].
\end{equation}
Similarly, the volume filling factor of $N$ wakes is given by
\begin{equation}\label{fV}
f_V\simeq \frac{9}{20} N \left(\frac{R_0}{r}\right)^3
        \left[\left(\frac{X}{R_0}\right)^{5/3} -1\right].
\end{equation}
Both estimates assume that the wakes do not
intersect in three dimensions, which makes them slight
overestimates.

The r.m.s.\  turbulent velocity averaged over the wake length $X$ is given by
\begin{eqnarray}\label{v0w}
v_0&=&\left(X^{-1}\int_0^X v_{0x}^2\,dx\right)^{1/2}\nonumber\\
&\simeq& \sqrt3 V\ini\left\{\frac{R_0}{X}
                \left[1-\left(\frac{R_0}{X}\right)^{1/3}\right]\right\}^{1/2}\nonumber\\
        &\simeq&\sqrt3 V\ini\left(\frac{\Reycr}{\Rey\ini}\right)^{3/2}
                \left(1-\frac{\Reycr}{\Rey\ini}\right)^{1/2}\;.
\end{eqnarray}
Similarly, the typical turbulent scale can be identified with the average
wake width,
\begin{equation}\label{l0wake}
l_0\simeq\frac34 R_0 (X/R_0)^{1/3}\;.
\end{equation}

The area covering factor and the volume filling factor of the wakes sensitively
depend on the subcluster mass $m$ and are larger for larger $m$
(which is consistent with the idea that turbulence produced in major merger
events fills the volume).
We also note the strong dependence of the covering and filling factors on
the Reynolds number:
$f_S\propto\Rey^4$ and $f_V\propto\Rey^5$ (for $X/R_0 \gg 1$).

It is clear from Eqs~(\ref{fS}) and (\ref{fV}) that one can have $f_V\ll1$ but
$f_S\simeq1$ (note that $R_0\ll r$). When this is the case, the wakes are
widely separated in space, but their images overlap in the two-dimensional
projection onto the sky plane.

\subsubsection{Subcluster wakes}\label{SW}
For subclusters of a mass $m=3 \times 10^{13}M_\odot$,
we adopt $R_0=100\kpc$,
$V\ini=\cs$ and $\lambda=1\kpc$ to obtain, from Eq.~(\ref{XR0a}), an
estimate
\begin{equation}\label{low}
\frac{X}{R_0} \simeq 27 \left(\frac{R_0}{100\kpc}\right)^3
\left(\frac{4\lambda\delta}{1\kpc}\right)^{-3}
\left(\frac{\Reycr}{400}\right)^{-3}\;.
\end{equation}
With the scaling of Eq.~(\ref{mergerrate}), the merger rate of subclusters
of this mass
is about 5 times larger than that of major mergers; thus, we assume that
$N=5$ subclusters of this mass can (almost) simultaneously fall into a
larger cluster.
The area covering and volume filling factors of $N=5$ wakes
within the radius $r=r_\mathrm{vir}\approx3\Mpc$ are estimated as
\[
f_S\simeq 0.2\,\xi\,\frac{N}{5}\,\left(\frac{R_0}{100\kpc}\right)^6
\left(\frac{\Reycr}{400}\right)^{-4}
\left(\frac{4\delta\lambda}{1\kpc}\right)^{-4}\;,
\]
and
\[
f_V\simeq 0.02\,
\frac{N}{5}\,\left(\frac{R_0}{100\kpc}\right)^8
\left(\frac{\Reycr}{400}\right)^{-5}
\left(\frac{4\delta\lambda}{1 {\rm kpc}}\right)^{-5}
\;.
\]

The covering and filling factors strongly depend on the poorly known
viscosity, parameterized with $\delta$. For $\delta\la0.16$, we obtain
$f_S\ga1$, but the volume filling factor remains smaller than unity for
$\delta\ga0.1$. Furthermore, both $f_S$ and $f_V$ depend on high powers of
another poorly known parameter, the stripping radius $R_0$. Hence, properties
of the subcluster wakes can be rather different in apparently similar
clusters. In addition, results of numerical simulations of turbulent wakes
should be treated with caution as otherwise reasonable approximations,
numerical resolution, and numerical viscosities can strongly affect the
results.

Upper limits on the covering and filling factors follow if we assume that
the wake length is equal to or exceeds the region size, $X=2r_\mathrm{vir}=6\Mpc$, which is
obtained in Eq.~(\ref{XR0a}) if $\Rey\ini/\Reycr>4$ (and $\lambda=1\kpc$):
\[
f_S\la 0.5(4\delta)^{-4}\;, \qquad f_V\la 0.08(4\delta)^{-5}\;,
\]
which yields $f_S\la2$ and $f_V\la0.5$ for $\delta\ga0.2$.

Thus, wakes from subclusters of a mass $3 \times 10^{13}M_\odot$
can occupy just a small fraction
of the total volume within the viral radius of a cluster, but their
area covering factor can be substantial.
Given the sensitive dependence on the poorly known value of the
Reynolds number, it appears reasonable to assume that
$f_S=O(1)$, that is any line of sight within
the virial radius will have good chance to
intersect at least one turbulent wake.

The r.m.s.\  turbulent velocity averaged over the wake length follows from
Eq.~(\ref{v0w}) as $v_0\simeq260\kms$ if averaged along the whole
length $X\simeq2.7\Mpc$, and $v_0\simeq190\kms$ within the cluster
virial radius (with
$X=2r_\mathrm{vir}\simeq6\Mpc$).
 The average turbulent
scale follows from Eq.~(\ref{l0wake}) as $l_0\simeq200\kpc$.

Takizawa (2005) has
recently studied turbulence generated by a subcluster
of total mass of $10^{14}M_\odot$,
gas core radius of $100\kpc$
and a much higher central density $\simeq 3 \times 10^{-2}\cm^{-3}$,
moving through a
uniform medium about 100 times less dense. Turbulent velocities
obtained in those simulations, 300--$500\kms$
(see also Norman \& Bryan 1999; Ricker \& Sarazin
2001; Schueker et al.\ 2004), are in a reasonable agreement with our estimates.

\subsubsection{Galactic wakes}\label{GW}
The stripping radius of galaxies could be estimated similarly to that of subclusters,
but the arguments are complicated by the replenishment of interstellar gas
by stellar winds, magnetic fields that affect the Kelvin--Helmholtz
instability, etc. Both numerical models (Portnoy et al.\ 1993; Balsara et al.\ 1994;
Acreman et al.\ 2003) and observations (Sun et al.\ 2005)
indicate that gas within
\begin{equation}\label{stripgal}
R_0=3\mbox{--}5\kpc
\end{equation}
of the centre of a massive elliptical
galaxy can remain unstripped. We assume that the Reynolds number based on this
scale,  $\Rey\simeq(10$--$15)\delta^{-1}$, is large enough to produce a
turbulent wake, e.g., because $\delta$ is small enough.

Consider a rich galaxy cluster, where $N\approx100$
galaxies are found within the gas core radius
$r\cl =180\kpc$ (Sarazin 1988). From Eq.~(\ref{fS}), the
area covering factor of galactic wakes in this region is unity if
\begin{equation}\label{XR0}
\frac{X}{R_0}\simeq 30\mbox{--}15\;,
\quad
X\simeq 100\mbox{--}70\kpc\;,
\end{equation}
where the range corresponds to that in Eq.~(\ref{stripgal}).
Wakes of this length would require $\Rey\ini/\Reycr\simeq3$,
which is obtained, e.g., for $\delta\simeq0.01$ if $\Reycr=400$.

The r.m.s.\ turbulent velocity and scale averaged along the wake follow from
Eqs~(\ref{v0w}) and (\ref{XR0}) as
\begin{equation}\label{v0gal}
v_0\simeq300\kms\;,\quad
l_0\simeq8\kpc
\end{equation}
for $R_0=4\kpc$ and $X/R_0=20$.
The volume filling factor of such wakes is $f_V\simeq0.07$.

The size of galactic wakes required to cover the projected cluster area,
given by Eq.~(\ref{XR0}), does
not seem to be unrealistic. For example, Sakelliou et al.\ (2005) have
observed a wake behind a massive elliptic galaxy (mass of order
$2\times10^{12}M_\odot$) moving through the intracluster gas at a speed about
$v_\mathrm{c}\simeq1000\kms$. The length of the detectable wake is about
$X\simeq130\kpc$ (assuming that it lies in the sky plane), and its mean radius
is $40\kpc$ (obtained from the quoted volume of about $2\times10^{6}\kpc^3$).
These authors argue that the wake is produced by the ram pressure stripping of
the interstellar gas. The projected area of the wake is about $10^4\kpc^2$, as
compared to $10^3\kpc^2$ for the wake parameters derived above.
This wake has been detected only because it is exceptionally strong, and it is
not implausible that weaker but more numerous galactic wakes can cover the
area of the central parts of galaxy clusters.

We conclude that subcluster wakes are likely to be turbulent,
but galactic wakes can be laminar if the viscosity of the intracluster gas is
as large as Spitzer's value. Given the uncertainty of the physical nature (and
hence, estimates) of the viscosity of the magnetized intracluster plasma,
we suggest that turbulent galactic wakes remain a viable
possibility. Both types of wake have low volume filling factor but can have an area
covering factor of order unity.

\section{Magnetic field in the intracluster gas}\label{MFDT}

In this section we discuss the amplification of an initially weak seed
magnetic field by the fluctuation dynamo operating in the intracluster
gas. The seed field itself can be produced by a wide range of
mechanisms (Appendix~\ref{seed}; see also Ruzmaikin et al.\ 1989; Widrow 2002;
Brandenburg \& Subramanian 2005).
We first discuss the fluctuation dynamo in general terms.
These general ideas are then applied to the various contexts
of intracluster turbulence discussed above. First, we consider
the merger epoch when the turbulence can be assumed to be
in a statistically steady state, then the later epochs after the
driving by the merger had ceased and the turbulence decays,
and finally to magnetic field generation in turbulent wakes.

\subsection{The fluctuation dynamo}\label{TFD}
The evolution of a magnetic field embedded into a flow of conducting fluid is
controlled by the
magnetic Reynolds number defined, similarly to Eq.~(\ref{Reyn}), as
\[
\Rm=\frac{v_0l_0}{\eta}\;,
\]
where $\eta$ is the magnetic diffusivity (inversely proportional to the
electric conductivity).

The exponentially fast amplification of an initially weak magnetic field by
a random flow (called the
{\it fluctuation dynamo\/}) is a result of a random stretching of  magnetic field by
the local velocity shear (see reviews in Zeldovich et al.\ 1990 and Brandenburg \&
Subramanian 2005). For $\Rm\gg1$, magnetic field is nearly frozen into the
flow. Then, due to the random stretching, magnetic field lines grow longer,
that is $B/\rho$ increases, where $\rho$ is the gas density.
For flows with $\rho$ approximately constant, the
magnetic field will be amplified. Such amplification comes at the cost of a
decrease in the scale of field structures in the directions perpendicular to
the stretching (i.e., on average
in all directions if the flow is statistically isotropic).
This enhances Ohmic dissipation and the latter ensures that the correlation function
of magnetic field can grow exponentially as an eigenfunction if the Lorentz force is
negligible (the kinematic dynamo). The growth occurs
under a fairly weak condition $\Rm > \Rmcr \simeq 30$--100 (where the
variation within the range depends on the form of the velocity
correlation function).
If $v_l$ is the velocity at a scale $l$, the $e$-folding time for the magnetic
field is roughly equal to the eddy turnover time $l/v_l$. In the
Kolmogorov turbulence, where $v_l\propto l^{1/3}$, the $e$-folding time is shorter at
smaller scales, $l/v_l\propto l^{2/3}$, and so smaller eddies amplify the field
faster.

Since $\eta\ll\nu$ in the rarefied intracluster plasma (e.g., Brandenburg \&
Subramanian 2005), we have
$\Rm\gg\Rey$. Therefore, $\Rm\gg\Rmcr$ if  $\Rey\ga100$, so that random
motions in galaxy clusters will be a dynamo
for any Reynolds number which is large
enough to make them turbulent.

Numerical simulations of magnetic field evolution in turbulent flows
confirm that the fluctuation dynamo action readily occurs in forced and
convective turbulent flows (Meneguzzi, Frisch \& Pouquet 1981;
Cattaneo 1999;
Haugen et al.\ 2003, 2004; Maron, Cowley \& McWilliams 2004;
Schekochihin et al.\ 2004), especially when $\Rm
\ge \Rey$. Such simulations are also able to follow the fluctuation dynamo
into the non-linear regime where the Lorentz forces becomes strong enough to
affect the flow as to
saturate the growth of magnetic field.

In the kinematic regime,
the field is predicted to be intermittent, i.e., concentrated into
structures whose size, in at least one dimension, is as small as the
resistive scale
\begin{equation}\label{leta}
l_\eta=l_0 \Rm^{-1/2}
\end{equation}
in a single-scale flow
(e.g., Ruzmaikin et al.\ 1989; Zeldovich et al.\ 1990).
We emphasize that magnetic field at the small Ohmic diffusion scale is
produced by the shear of the flow at a larger scale $l_0$.

In a turbulent
flow, where a broad spectrum of motions is present, flow at each scale $l$
would produce magnetic
structures at all scales down to
the corresponding Ohmic scale.
In the kinematic regime this would correspond to a set of
eigenfunctions, each with
a distinct growth rate
 $v_l/l$. The fastest
growing
eigenfunction is due to stretching by the
smallest eddies with $\Rm(l) > \Rmcr$, where $\Rm(l) = \Rm (l/l_0)^{3/4}$.
These are the viscous scale eddies, with
$l = l_\nu = l_0 \Rey^{-3/4}$,
provided $\Rm/\Rey > \Rmcr$. However in the nonlinear
regime, when the
fastest growing mode saturates, larger scale modes could still grow.
Since most
of the kinetic energy is contained at the scale $l_0$, the dominant magnetic
scale could still be determined by dynamo action due
to eddies of scale $l_0$ and,
especially, by the subtle details of the dynamo saturation.
We now discuss how the dynamo action could saturate.

Nonlinear effects can modify the resulting magnetic structures,
although it is as yet not clear
in what way (cf.\ Haugen et al.\ 2003, 2004; Schekochihin et al.\ 2004). A
simple model of Subramanian (1999) suggests that the
smallest scale of the magnetic structures will be renormalized in the
saturated state to become
\begin{equation} \label{lB}
l_B\simeq l_0\Rmcr^{-1/2}\;,
\end{equation}
instead of the resistive scale $l_\eta$.
This
essentially happens via a renormalization
of the effective magnetic diffusivity in the models of
Subramanian (1999, 2003).
In other words, it is
suggested that the dynamo action can be saturated via a reduction of the
effective magnetic Reynolds number down to its critical value for the dynamo
action.
\footnote{Saturation could also happen if
the stretching properties of the flow
are suppressed by the Lorentz force (Kim 1999).
This can perhaps be described as a reduction of the
effective magnetic Reynolds number
to its critical value
$\Rmcr$ of the kinematic dynamo, a feature which however has not
yet been studied in the model of Kim (1999)}.

Such results, however plausible they are, require further
substantiation, e.g., by numerical simulations.
Dynamo simulations
of Haugen et al.\ (2003, 2004) with
$\nu/\eta=1$, where $\Rmcr\approx35$, show that
magnetic energy per unit logarithmic interval of $k$, $kM_k$, has
a broad maximum at $k\sim 9$, a scale about 6 times smaller
than the forcing scale, but
a factor of about four
larger than the resistive scale
given by
 $k_\eta=k_0\Rm^{1/2}\simeq(20$--$30)k_0$ for $\Rm=420$--$960$,
in agreement with the
above idea and Eq.~(\ref{lB}).
(We note, however, that it is not quite clear how significant is the difference
or agreement here since all the estimates have factors of order unity omitted, which
can be important at the modest values of $\Rm$ available.)
Further, these simulations also show that
the value of $k_B$ does
not scale with $\Rm$ when $\Rm$ is increased from about
$420$ to $960$, confirming Eq.~(\ref{lB}).
However, the magnetic spectrum is rather broad and it is difficult to identify
accurately the dominant magnetic scale in those simulations. Nevertheless,
it is clear that the nonlinear magnetic field distribution is
less intermittent (i.e., its scale is larger) than at the kinematic stage.
Below we present evidence for this nonlinear behaviour in our own
simulations of the fluctuation dynamo.
The values of magnetic Reynolds number accessible now in such
simulations are too modest to make any confident conclusions,
but we believe that our approach to the saturation of the fluctuation
dynamo is consistent with the evidence available.

For the Kolmogorov turbulence, in the kinematic regime,
the corresponding resistive scale for the marginal mode is predicted to be
$l_\eta \simeq l_0 R_{\rm m}^{-3/4}$ (Subramanian 1997; Brandenburg \&
Subramanian 2005), when it is larger than the viscous cut-off scale
(that is when $l_\eta > l_\nu = l_0\Rey^{-3/4}$ or $\Pm < 1$).
This scaling is different from that
 in Eqs~(\ref{leta}) and (\ref{lB}), which apply to
a single-scale flow, because the shearing rate now is
$(v_0/l_0)(l/l_0)^{-2/3}$ at any scale $l$ in the inertial range.
For the marginal magnetic mode (neither growing nor decaying),
the shearing
 is balanced
by dissipation at $l=l_\eta$ which occurs at a rate $\eta/l_\eta^2$.
In the
intracluster gas, we generally have $l_\eta < l_\nu$. However
in the saturated state one may have $l_B > l_\nu$ for $\Rey > \Rmcr$.
This would then suggest a different scaling, $l_B\simeq l_0\Rmcr^{-3/4}$
instead of Eq~(\ref{lB}).
Nevertheless,
in both the simulations presented
below and in real clusters the flow is not strongly turbulent,
i.e., it has no extended Kolmogorov inertial range
because the Reynolds number is not very large; so
Eq.~(\ref{lB}) can remain a better approximation.
There is some evidence for this from the
simulations, in that Eq.~(\ref{lB}) agrees better
with the wave number at which the magnetic spectrum peaks.
We shall therefore use Eq.~(\ref{lB}) in our estimates.

We note that properties of MHD turbulence can
depend on the ratio $\Pm=\nu/\eta$, known as the magnetic Prandtl
number. The intracluster gas has $\Pm\gg1$ if Spitzer's viscosity and
resistivity are adopted; realistically large values of
$\Pm$ are not accessible to computer simulations, but we shall discuss
simulations with a modestly large value of $\Pm$ in what follows.

\subsection{Application to cluster turbulence}\label{FDDT}
Here we present semi-quantitative estimates to characterize the fluctuation dynamo
in the intracluster gas, before discussing, in Sect.~\ref{SFD}, direct numerical
simulations of the fluctuation dynamo.

\subsubsection{The epoch of cluster formation}

During the epoch of major mergers, we expect that the intracluster medium
is involved in a steady-state, driven turbulence, for which we assume the
Kolmogorov spectrum. Due to the action of eddies
at scale $l$, the r.m.s.\ magnetic field grows exponentially at a rate
\begin{equation}
\gamma(l) =\frac{v_l\ini}{l}
= \frac{1}{t_0\ini}\left(\frac{l}{l_0\ini}\right)^{-2/3}\;,
\qquad
t_0\ini\simeq \frac{l_0\ini}{v_0\ini}\;,
\label{gammerge}
\end{equation}
where subscript `i' refers to the initial, steady state of the intracluster turbulence.
As summarized in the last paragraph of Sect.~\ref{TPDCF}, the turbulent speed
and scale can be adopted as
$v_0\ini=300\kms$ and $l_0\ini=150\kpc$, respectively. Assuming that the
 driven turbulence lasts for $t\f=3\Gyr$,
 we obtain an amplification exponent of the magnetic field
$\Gamma = \gamma(l) t\f\approx6$
 at $l=l_0\ini$ (and larger at smaller scales).
 So the seed field can be amplified by a
factor $400$ by motions
at $l=l_0\ini$ during this time
(the amplification factor is larger at smaller scales).
For a seed magnetic field of
$10^{-8}\G$, this amplification is
sufficient to explain the observed magnetic fields; this implies that
the observed magnetic fields are plausibly in the saturated state and the
Lorentz force can now
affect significantly the velocity field in galaxy clusters.

\subsubsection{The epoch of decaying turbulence}

After the driving forces have been diminished, the turbulence decays.
Both the instantaneous outer scale and the viscous scale
then increase with time (the latter, because of the decrease
in the Reynolds number). In this situation it is more useful
to estimate the growth rate at a fixed scale $l$, which
belongs to the turbulent spectrum for a long time
and can hence lead to a fluctuation dynamo,
instead of considering an evolving viscous scale (where the dynamo time scale
is the shortest).
The instantaneous growth rate of the r.m.s.\ magnetic field due to motions
at a fixed scale $l$ decreases with time as
\begin{eqnarray}
\gamma(l,t)&=&\frac{v_l(t)}{l}\nonumber\\
&=&\frac{v_l\ini}{(l^2l_0\ini)^{1/3}}
\left(\frac{t-t\f}{t_0\ini}\right)^{-\left(\frac{1}{2}\alpha+\frac{1}{3}\beta\right)},
\quad t\gg t_0\ini\;. \label{gammalt}
\end{eqnarray}
Here we have adopted a Kolmogorov spectrum with
$v_l=v_0(t)[l/l_0(t)]^{1/3}$ and for numerical estimates
take $\alpha=6/5$ and $\beta =2/5$.
For a growth rate evolving with $t$, magnetic field evolves as
$B\propto\exp{\int_{0}^t\gamma(l,t')\,dt'}$. If
turbulence is maintained in a steady state during an initial
period $t\leq t\f$ and
then $\gamma$ decreases as in Eq.~(\ref{gammalt}),
the amplification exponent
for the Kolmogorov spectrum follows as
\begin{equation}\label{IntGamma}
\int_{0}^t\gamma(l,t')\,dt'=\Gamma(t)
\left(\frac{l}{l_0\ini}\right)^{-2/3}\;,
\end{equation}
where
\[
\Gamma(t)=
\left\{
\begin{array}{ll}
\displaystyle
\frac{t}{t_0\ini}\;,            &t\leq t\f+t_0\ini\;,\\[7pt]
\displaystyle
1+\frac{t\f}{t_0\ini}+\frac{1}{\zeta}
        \left[\left(\frac{t-t\f}{t_0\ini}\right)^{\zeta}-1\right],
                                                &t>t\f+t_0\ini\;,
\end{array}
\right.
\]
where $\zeta=1-\frac{1}{2}\alpha-\frac{1}{3}\beta$.
We have applied the power-law
(\ref{gammalt}) only at $t>t\f+t_0\ini$ and assumed that the decay does not
affect the growth rate before that time.
Assuming that the
turbulence starts decaying after a time $t\f=3\Gyr$, we
obtain, at $t=5\Gyr$, that the energy-range speed reduces down to
$v_0\approx130\kms$, whereas the energy-range scale increases to
$l_0\approx260\kpc$.
The amplification exponent of magnetic field due to motions
at $l=l_0\ini$ is obtained as $\Gamma\approx9$
at $t=5\Gyr$ (consisting of 7 at $t=t\f+t_0\ini=3.5\Gyr$ and
only less than 2 at later times),
so that the seed field could be amplified by a factor
about $6\times10^3$ by the end of the decay phase if
it were too weak to bring the dynamo to the saturated
state earlier. In order to obtain magnetic field of $1\mkG$
at this scale for $t=5\Gyr$, a seed field of
$2 \times 10^{-10}\G$ would be sufficient.
Again the amplification is larger  due to smaller scale eddies which remain
part of the turbulent cascade as the turbulence decays.

\subsubsection{Dynamo action in wakes}
Using the turbulent speed and scale averaged over the wake length, as derived
in Sect.~\ref{CFTW}, we obtain
magnetic field growth time scales
$\gamma^{-1}\simeq\l_0/v_0\simeq 0.8\Gyr$ for
subcluster wakes and $3\times10^7\yr$ for galactic wakes. At a given position,
time available for the dynamo action is $X/V\ini$, where $V\ini\simeq1000\kms$
is the speed of a subcluster or a galaxy. Therefore, the dynamo amplification
exponent is given by
\[
\Gamma  \simeq \frac{v_0}{V\ini}\,\frac{X}{l_0}
        \simeq3
\]
for both subclusters and galaxies, which implies additional amplification by a
factor 20 at the outer scale.

\subsubsection{Magnetic field strength in the intracluster gas}\label{MFSIG}

The maximum local magnetic field strength produced by the turbulent dynamo will
be, presumably, close to equipartition with the turbulent energy:
\begin{eqnarray}\label{Beq}
B_\mathrm{eq}&=&(4\pi\rho v_0^2)^{1/2}\nonumber\\
             &\simeq&  3\mkG
\left(\frac{n}{10^{-3}\cm^{-3}}\right)^{1/2}
\!\!\left(
\frac{v_0}{200\kms}
\right),
\end{eqnarray}
where $\rho$ is the gas density.
[We note that some models of nonlinear fluctuation dynamo predict stronger local
magnetic fields (Belyanin et al.\ 1993, 1994), but here we adopt a conservative
limit (\ref{Beq}).]  As discussed in Sect.~\ref{TFD}, magnetic
field produced by the fluctuation dynamo is expected to be spatially
intermittent (especially at early stages of dynamo action),
i.e., represented by intense filaments and sheets whose volume
filling factor is less than unity. Numerical simulations
and analytical models recently reviewed by Brandenburg \& Subramanian (2005)
suggest that magnetic
sheets and ribbons are prevalent, whose thickness is given by Eq.~(\ref{lB}) and
whose other two dimensions are of the order of the turbulent scale $l_0$
(see Fig.~\ref{snapshots}). Then the volume filling
factor of magnetic structures {\it within a single turbulent cell\/}
in the statistically steady state can be estimated as
\[
f_B=\frac{l_B l_0^2}{l_0^3}\simeq\Rmcr^{-1/2}\approx0.17\;,
\]
where the numerical value refers to $\Rmcr=35$.
Therefore, the r.m.s.\ magnetic field within a turbulent cell is of order
\begin{eqnarray}\label{Brms}
\mean{B^2}^{1/2}&\simeq& f_B^{1/2}B_\mathrm{eq}\nonumber\\
        &\simeq&1.2\mkG\left(\frac{n}{10^{-3}\cm^{-3}}\right)^{1/2}
\left(\frac{v_0}{200\kms}\right)\nonumber\\
        &&\mbox{}\times \left(\frac{\Rmcr}{35}\right)^{-1/4},
\end{eqnarray}
which implies that magnetic energy density in a saturated dynamo state is a
factor $\Rmcr^{1/2}\simeq6$ times smaller than the turbulent energy density.
(Here and below angular brackets denote averaging.)
We emphasize that weaker volume-filling magnetic fields are also present; their
contribution to the magnetic energy density can be somewhat smaller than
that of the intermittent part. In agreement with this estimate, magnetic energy
density is about
0.2 (0.33) of the kinetic energy density in Model 1 (Model 2) of
the numerical simulations discussed in Sect.~\ref{SFD}.
As discussed by Haugen et al.\ (2004, their Fig.~14), this ratio weakly varies
with magnetic Reynolds number; in their simulations, the variation is from about
0.25 for $\Rm=420$ to 0.4 for $\Rm=960$, both with $\Pm=1$.
For $\Pm=30$, their simulations yield the magnetic-to-kinetic energy ratio of
about unity.
Thus, the flow at larger magnetic Prandtl number appears to be producing
more magnetic energy; perhaps magnetic structures fill turbulent cells more
densely, and/or a smoothly distributed magnetic field is stronger.
This feature is yet to be understood. What is, however, important for our
immediate purpose here, is that the fluctuation dynamo produces magnetic fields
whose energy density is comparable to the kinetic energy density of the turbulence.

If the volume filling factor of the turbulent flow $f_V$ is less than unity, as in the
case of turbulent wakes of subclusters and galaxies, the r.m.s.\ magnetic
field in the cluster volume is obtained from Eq.~(\ref{Brms}) by further
multiplication by a factor $f_V^{1/2}$;
this is a measure of the total magnetic energy of the cluster. However, this
quantity has little physical significance because it would not result from any
local magnetic measurement. In this sense, the local value (\ref{Brms}) is
more meaningful; it is presented in Table~\ref{res} together with other
quantities that characterize turbulence and magnetic fields at various stages
of the cluster evolution.

\begin{table*}
\begin{minipage}{135mm}
\caption[]{\label{res}Summary of turbulence and magnetic field parameters at
various stages of cluster evolution: duration of the stage (the last two
stages represent steady states), the r.m.s.\ velocity $v_0$ and scale $l_0$
of turbulence and eddy turnover time $t_0$ (for the decaying turbulence,
values for the middle of the decay stage are given, 2\,Gyr after its start),
the equipartition magnetic
field $B_\mathrm{eq}$
given by Eq.~(\ref{Beq}) (i.e., maximum field strength within a turbulent cell),
thickness of magnetic filaments and sheets $l_B$,
defined in Eq.~(\ref{lB}), for the statistically steady
state of the dynamo, the r.m.s.\
magnetic field within a turbulent cell $B_\mathrm{rms}\equiv\mean{B^2}^{1/2}$
given by Eq.~(\ref{Brms})
(the latter two obtained for $\Rmcr=35$), and
finally the standard deviation of the Faraday rotation measure $\sigma_\RM$
[calculated using Eq.~(\ref{sigmaRM})
 for the volume filling turbulence within $500\kpc$ of the centre
and path length of $750\kpc$ in the first two lines, and
assuming one transverse wake along the line of sight in the last two lines,
using Eq.~(\ref{sigmaRMw})].
Subcluster mass of $3\times10^{13}M_\odot$ has been assumed.}
\begin{tabular}{@{}lccccccccc@{}}
Evolution stage &Duration&$v_0$     &$l_0$    &$t_0$     &$B_\mathrm{eq}$&$l_B$   &$\mean{B^2}^{1/2}$ &$\sigma_\RM$ \\
                &[Gyr]   &[$\!\kms]$&[kpc]    &[Gyr]     &[$\mu$G]       &[kpc]   &[$\mu$G]               &[$\!\radm$] \\[7pt]
Major mergers   &4       &300       &150      &0.5\phzero&4              &25      &1.8                    &200\\
Decaying turbulence
                &5       &130       &260      &2.0\phzero&2              &44      &0.8                    &120\\
Subcluster wakes&        &260       &200      &0.8\phzero&4              &34      &1.6                    &110\\
Galactic wakes  &        &300       &\tphzero8&0.03      &4    &\phzero\tphzero1.4&1.6                    &\tphzero5\\
\end{tabular}
\end{minipage}
\end{table*}

\section{Simulations of the fluctuation dynamo}\label{SFD}

We have simulated the generation and subsequent
decay of dynamo-active turbulence using
the numerical model of the fluctuation dynamo by
Haugen et al.\ (2003, 2004), where isothermal, viscous,
electrically conducting, compressible gas is driven by a random force imposed as a
source in the Navier--Stokes equation. The Navier--Stokes, continuity and
induction equations are then solved in a Cartesian box of a size $D$ on a
cubic grid with $256^3$ mesh points. The driving force $\vect{f}$ is sinusoidal
in the spatial coordinates, transversal ($\vect{f}\perp\vect{k}$ with
$\vect{k}$ the wave vector of the force), and localized in the wave-number space
about a certain wave number $k=k\f$, so it drives almost
incompressible, vortical motions in a certain wavelength range around $2\pi/k\f$
(see Haugen et al.\ 2004 for details).
The direction of the
wave vector of the force and its phase change randomly every time step in the
simulations, so the force is effectively $\delta$-correlated in time.

\begin{figure}
\centerline{\includegraphics[bb=98 100 448 341,width=0.495\textwidth]{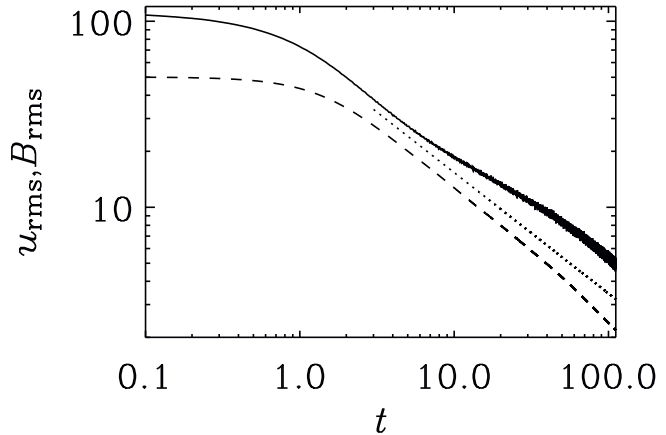}}
\caption{\label{decay}
The evolution of the r.m.s.\ fluid velocity (solid) and magnetic field
(dashed) in driven and then decaying turbulence with $\tilde{k}\f=5$
(Model 1) ($v_\mathrm{rms}\equiv \mean{v^2}^{1/2}$ and likewise for $B$).
Velocity is
measured in the units of diffusive speed at the driving wavelength,
$\nu k\f$, and magnetic field is expressed in similar velocity
units, $(4\pi\rho)^{1/2}\cs\nu k\f$.
Hence, $v_\mathrm{rms}$ numerically coincides with $\Rey$ in the statistically
steady state. Time is measured in the units of the initial
turnover time of the energy-containing eddies, $t_0\ini$.
Dotted line shows the asymptotics $(t-t\f)^{-0.65}$, where $t\f=0$
is the time when the forcing is turned off.
}
\end{figure}

We represent numerical results using the following units. (Tilde is used to
denote dimensionless quantities.) For a unit length
$d$, the computational domain size is equal to $D=2\pi d$. The wave number is
measured in the units of $d^{-1}$. In simulations with dimensionless forcing
wave number $\tilde{k}\f$, it is appropriate to adopt $k\f=2\pi/l_0$
for its dimensional value, where $l_0=150\kpc$ is
the turbulent scale in a merging cluster, as obtained in Sect.~\ref{TPDCF}.
Then the unit length is $d=\tilde{k}\f l_0/(2\pi)$ and the dimensional size
of the computational domain is $D=l_0\tilde{k}\f$. The unit density $\rho_0$
can be adopted to correspond to the number density of $n_0=10^{-3}\cm^{-3}$.
The unit speed is the speed of sound, $\cs=1000\kms$, so that the unit
magnetic field is $(4\pi\rho_0)^{1/2}\cs=15\mkG$.

Here we report results obtained with two values of the central driving wave number
$k\f$.
Some results were obtained with driving covering the range
of dimensionless wave numbers $\tilde{k}=4.5$--5.5, centred at
$\tilde{k}\f=5$ (Model~1). Results at
higher resolution (which was especially needed when $\Pm>1$), were obtained
with the driving wave-number range of $\tilde{k}=1$--2 centred at
$\tilde{k}\f=1.5$ (Model~2). In these latter runs, the computational box
contains just a few turbulent cells.

The intensity of the
driving was adjusted to obtain the r.m.s.\ Mach number
of the turbulence of about 0.1 which produces relative density
fluctuations of order $0.01$
(implying that only a small fraction of the total velocity is compressible).
The kinematic viscosity and magnetic diffusivity in most runs
are adopted to be equal to $\nu=\eta=2\times10^{-4}\cs d$ (i.e., a magnetic
Prandtl number of unity). This corresponds to $\Rey=\Rm\approx110$ in Model~1 and
$\Rey=\Rm\approx420$ in Model~2, which is
close to what is expected for $\Rey$ in the intracluster gas.
(We note, however, that the values of magnetic Reynolds number explored
here still are smaller than those expected in reality.)
We have also considered the case where magnetic diffusivity is 30 times smaller
than kinematic viscosity in runs with $\tilde{k}\f=1.5$,
$\nu=1.5\times10^{-3}\cs d$ and $\eta=\nu/30$, i.e.,
$\Pm=30$, $\Rey\approx44$
and $\Rm\approx1300$..
Results presented in what follows refer to the case $\Pm=1$ unless stated
otherwise.

In order to simulate dynamo action in forced and then decaying turbulence,
the flow had been driven until it
 reached a statistically steady state, with
a weak magnetic field introduced at the start of the simulation.
Then the system was evolved for some period (about
45 time units in Model~1),
 after which the driving force was switched off;
$t\f=0$ is the time when the driving halts.
The initial, weak magnetic field is random, with
energy density of about 0.6\% of the kinetic energy density in Model~1.

\begin{figure*}
\includegraphics[width=0.495\textwidth]{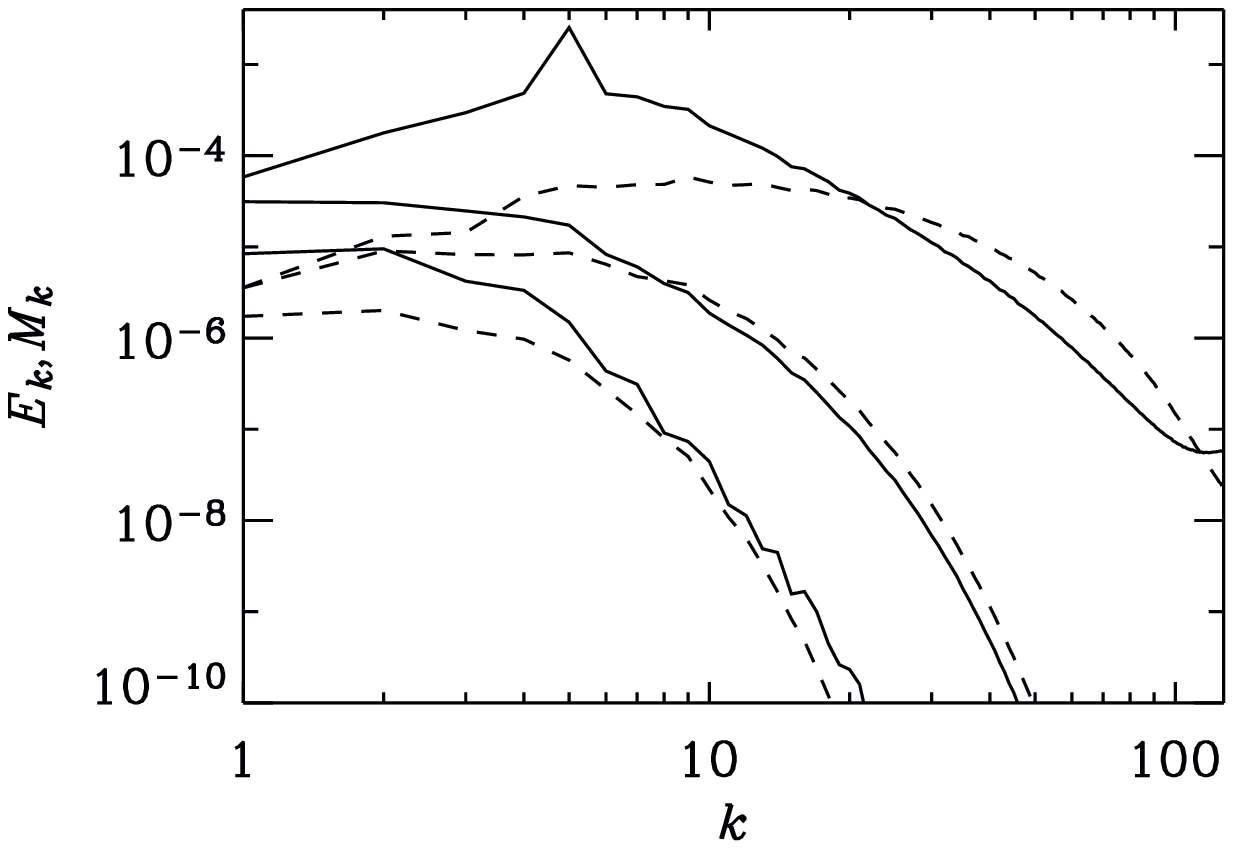}
\hfill
\includegraphics[width=0.495\textwidth]{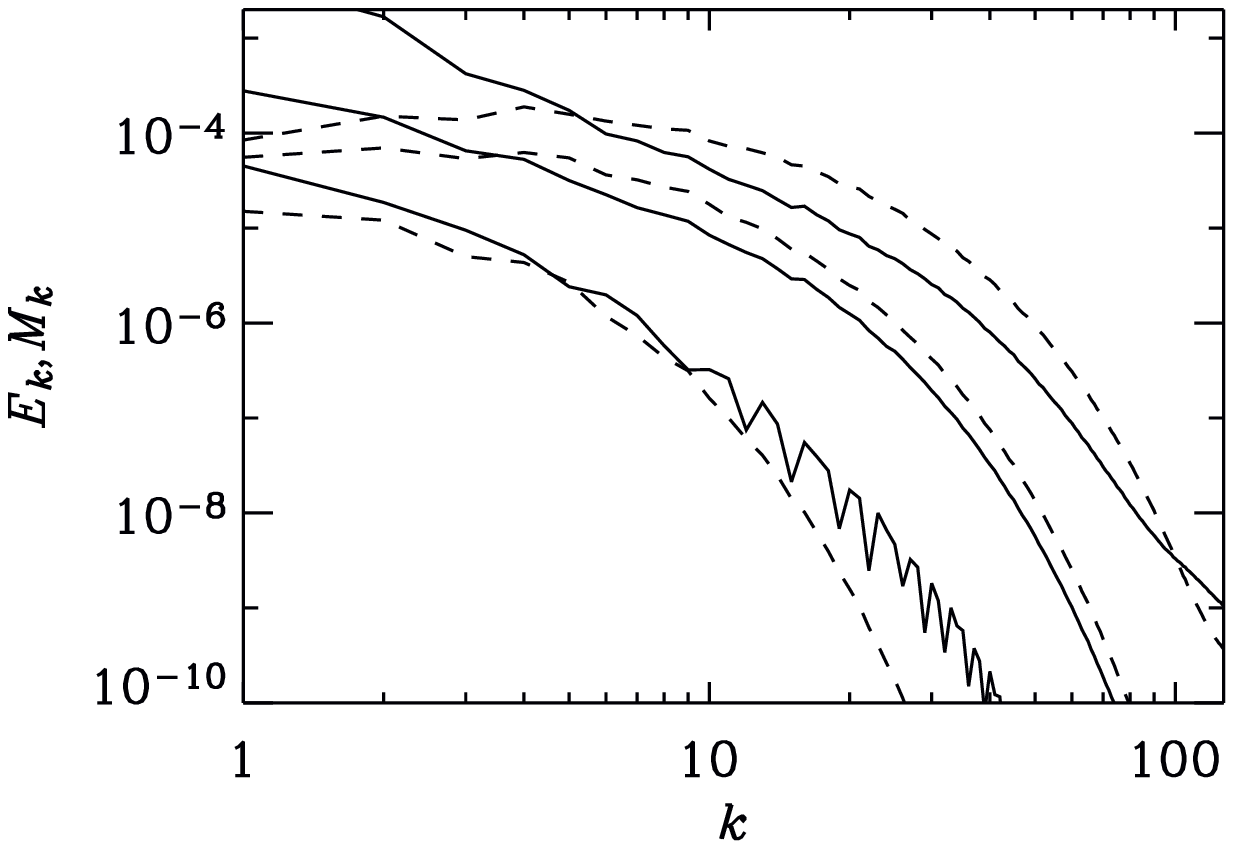}
\caption{\label{spectra}The energy spectra of kinetic (solid) and magnetic
(dashed) energies at various stages of evolution, $t/t_0\ini=0,\ 10,\ 50$,
with (a) $\tilde{k}\f=5$ -- Model 1, left panel, and (b) $\tilde{k}\f=1.5$ --
Model 2, right panel.
Spectra obtained at later times are at lower levels because of the decay of
turbulent energy.}
\end{figure*}

Figure~\ref{decay} shows the time evolution of
the suitably normalized r.m.s.\ velocity and magnetic field obtained
in Model~1, after the driving was switched off
[in fact, $v_\mathrm{rms}/(\nu k\f)$, shown with solid line,
is the Reynolds number based on the forcing scale,
with $v_0\approx v_\mathrm{rms}$].
The initial exponential growth of the r.m.s.\ magnetic field
that obtains (not shown in Figure~\ref{decay}),
is followed by its saturation at a level where its energy density is about
0.2 (0.3) of the turbulent energy density in Model 1 (Model 2).
More precisely, the r.m.s.\ values
of the turbulent velocity and magnetic field (measured in velocity units)
in the steady state of
Fig.~\ref{decay} are about $0.114\cs$ and $0.050\cs$, respectively, whereas the similar
quantities for $\tilde{k}\f=1.5$ are $0.116\cs$ and $0.065\cs$. The critical value of the
magnetic Reynolds number remains about 35 in both models.

The subsequent decay of both the
velocity and magnetic field strength can be approximated
by $(t-t\f)^{-0.65}$ for
$t\gg t_0\ini$, as shown with dotted line. This decay law is consistent with
Eq.~(\ref{Ekdecay}) which predicts the power law exponent of $-3/5$. However,
the alternative value $\alpha=10/7$ would result in the exponent of
$-5/7\approx-0.71$, which is also consistent with our numerical results.

With $l_0=150\kpc$ and $n=10^{-3}\cm^{-3}$, the r.m.s.\  turbulent velocity and
magnetic field strength in the steady state in Fig.~\ref{decay} are
$\mean{v^2}^{1/2}\approx 110\kms$ and $\mean{B^2}^{1/2}\approx 0.7\mkG$,
respectively. These results favourably agree with estimates presented in
Table~\ref{res} in the sense that in both cases the ratio of the turbulent and
Alfv\'en speeds is about 1/2, which confirms our estimate of the r.m.s.\ magnetic
field strength in Eq.~(\ref{Brms}).
In other words, if our simulations had stronger driving to achieve
$\mean{v^2}^{1/2}\approx 300\kms$, then the r.m.s.\ magnetic field would
be $\mean{B^2}^{1/2}\approx 2\mkG$, as in our analytical estimate.

The magnetic and kinetic energy spectra are shown
in Fig.~\ref{spectra}. In the statistically steady state (the upper
curves), kinetic energy in Model~1 peaks at $\tilde{k}\f=5$, the driving wave number.
However, magnetic energy has broad maximum at a significantly smaller
scale, apparently because of its intermittent structure. This difference
is better visible in the right panel that refers to Model 2 where we have
higher resolution. Similar simulations, but with a significantly higher
resolution $1024^3$ (Haugen et al.\ 2003, 2004), confirm that the magnetic energy
per unit logarithmic interval in the $k$-space, $kM_k$,
has a maximum at $k_B\approx6k\f$ in excellent agreement with Eq.~(\ref{lB})
with $\Rmcr=35$,
but no significant dependence of $k_B$ on $\Rm$ has been revealed.
The value of $\Rm$ is those simulations is about $1000$, so the deviation from the
scaling $k_B \sim k_\eta\propto\Rm^{-1/2}$ advocated by
Schekochihin et al.\ (2004) is by
a factor of 3--5 over the range $\Rm=400$--1000.

The length scales of the velocity
and magnetic fields can be characterized more precisely in terms of their
integral scales
\begin{equation}\label{iscale}
L_v=2\pi\frac{\int k^{-1}E_k\,dk}{\int E_k\,dk}\;,
\end{equation}
and similarly for the magnetic scale $L_B$; here
integration extends over the whole interval of $k$ available.
These scales are simply related to the longitudinal $l_\mathrm{L}$ and transverse
$l_\mathrm{N}$ integral scales of the magnetic fields by
$l_\mathrm{L}=\frac12 l_\mathrm{N}=\frac38L_B$ [Eq.~(12.91) of Monin \& Yaglom 1975],
and similarly for $L_v$, but only approximately because $\vect{v}$ is not solenoidal.
The dimensionless value of the longitudinal integral scale of magnetic field
in the steady state is then
obtained from Fig.~\ref{scale} as
$l_{\mathrm{B}}=\sqrt{2/\pi}l_\mathrm{L}\approx0.16$ (see Appendix~\ref{sRM}).
This agrees reasonably
well with the prediction from our heuristic estimates,
$l_B \simeq (2\pi/\tilde k_F) \Rmcr^{-1/2} \simeq 0.2$.
The time variation of these scales is shown in Fig.~\ref{scale}.
When the turbulence decays,
the integral scales of both velocity and magnetic field exhibit power-law
increase, in agreement with Eq.~(\ref{Ekdecay}); the growth slows down when
$L_v$ has grown to become comparable with the box size.

\begin{figure}
\centerline{\includegraphics[width=0.495\textwidth]{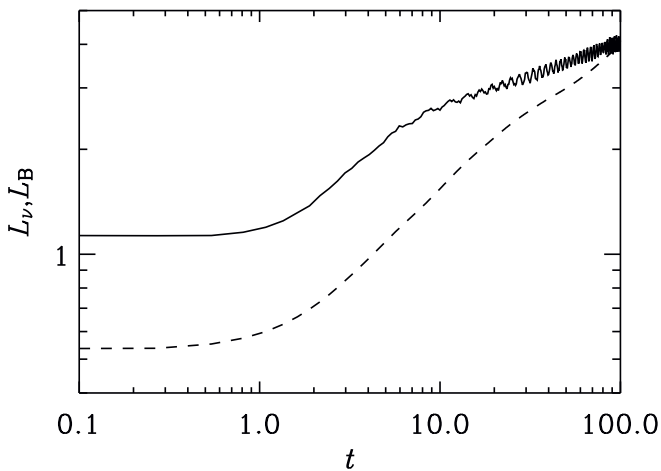}}
\caption{\label{scale}
Evolution of the integral scales of the velocity ($L_v$, solid line)
and magnetic fields ($L_B$, dashed line), as defined in Eq.~(\ref{iscale}),
with $\tilde{k}\f=5$ (Model~1).}
\end{figure}

Magnetic energy at small scales has, at
early times, excess over kinetic energy because magnetic field is very
intermittent, which is especially clearly visible in the right panel of Fig.~\ref{spectra}.
At later stages, magnetic field distribution becomes more homogeneous
and this feature disappears. Simultaneously, the scale of magnetic field
increases and becomes comparable to that of the flow, which is not the case at
early stages.

\begin{figure*}
{\includegraphics[width=0.485\textwidth,clip]{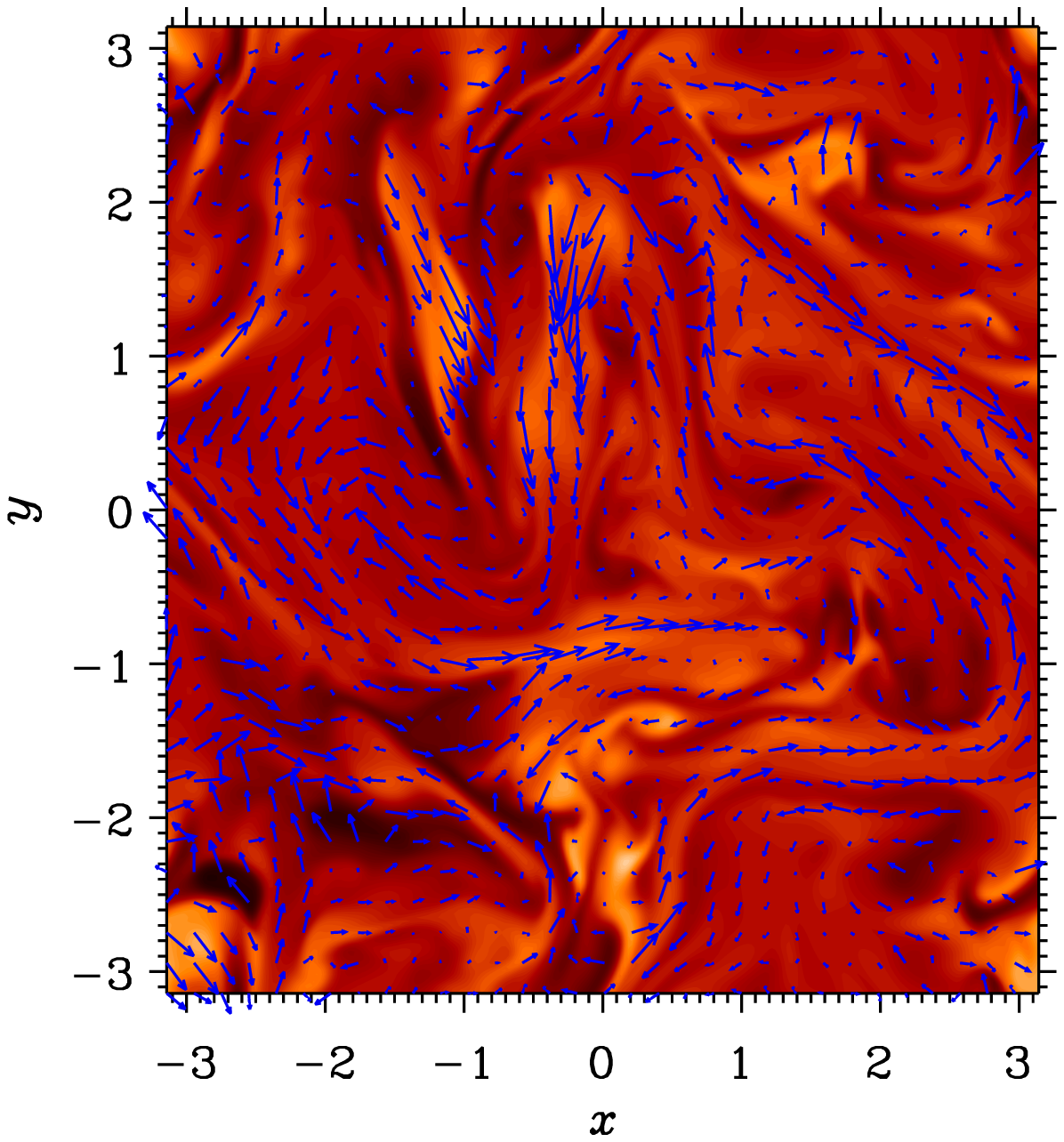}}
\hfill
{\includegraphics[width=0.5\textwidth,clip]{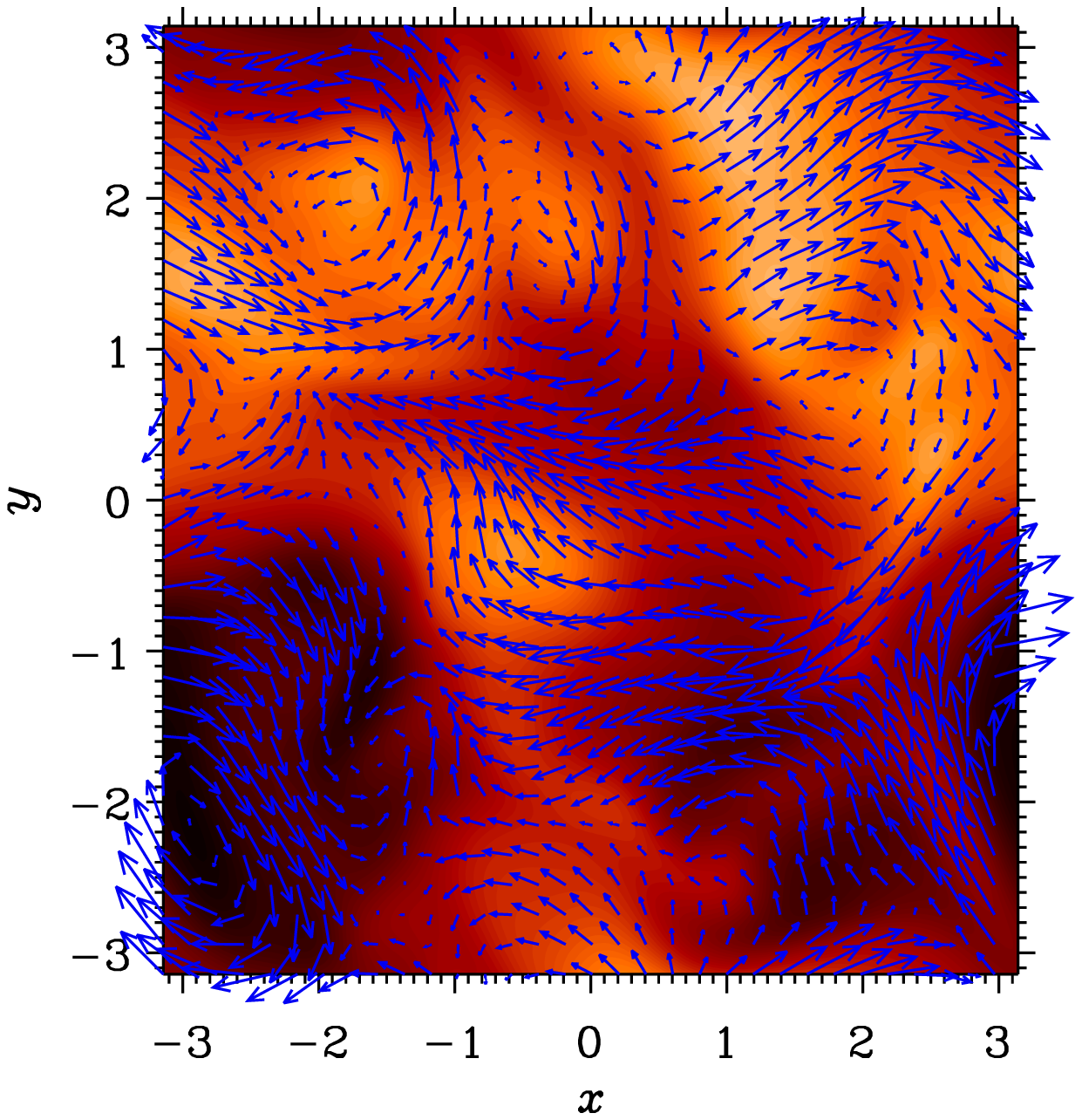}}
\caption{\label{snapshots}\label{snapshots1}
Snapshots of magnetic field in a cross-section through the middle of the
computational domain
of Model~2 at $t/t_0\ini=0.30$ when the system is in a statistically
steady state (left panel), and $t/t_0\ini=59.34$, at a late stage of decay
(right panel).
Here $\tilde{k}\f=1.5$, so each frame contains a few turbulent cells.
The magnitude of the field component perpendicular to the
plane of the figure is shown color coded (in shades of grey) with black
corresponding to field pointing into the
figure plane, and
lighter shades, to field pointing out of the plane. The field in the plane of
the figure is shown with vectors whose length is proportional to the field
strength.}
\end{figure*}

Figure~\ref{snapshots} illustrates (using Model 2)
the structure of magnetic field
in a turbulent flow in a
statistically steady state (left panel) and at a late stage of decay (right
panel).
The magnetic field produced by the fluctuation dynamo consists of an
intermittent part, represented by randomly distributed, intense magnetic
ribbons, sheets and filaments (which can even be folded), immersed in a sea of
volume-filling random magnetic field.
The intermittency gradually reduces as the turbulence decays together with
magnetic field because structures of smaller scale decay faster,
and the volume filling factor of magnetic field increases with time -- this
tendency can easily be seen in the right panel of Fig.~\ref{snapshots1}.

\section{Observational diagnostics}\label{OD}

\subsection{The Faraday rotation measure}\label{ODFRM}
An important observational diagnostic of the intracluster magnetic field is the
Faraday rotation of polarized radio emission of background sources (located
beyond the cluster or in its centre) produced in the intracluster gas. The
Faraday rotation is quantified by the Faraday rotation measure
\begin{equation}\label{RMdef}
\RM=K\int_L \nel\vect{B}\cdot d\vect{l}\;,
\end{equation}
where $\nel$ is the number density of free thermal electrons, the integral
is taken along the path length $L$ from the source to the observer, and
$K=0.81\FRM\cm^3\mkG^{-1}\p^{-1}$. For a magnetic field with zero mean value
$\mean{\vect{B}}=0$, the mean value of $\RM$ vanishes, whereas its standard
deviation can be represented in the form
(Appendix~\ref{sRM}; see also Burn 1966; Sokoloff et al.\ 1998),
\begin{equation}\label{sRMN}
\sigma_\RM\simeq\RM_0\sqrt{N}\;,
\end{equation}
where $\RM_0$ is the Faraday rotation measure produced in a single turbulent
cell of a size $l_0$ and $N=L/l_0$ is the number of the cells along the line
of sight.

Suppose that each turbulent cell contains randomly oriented magnetic sheets of
thickness $l_B$ where magnetic field strength is equal to $B_\mathrm{eq}$,
with a covering factor of order unity, as described in Sect.~\ref{MFSIG}.
Then the Faraday rotation measure produced in a single turbulent cell follows as
$\RM_0\simeq K\nel B_\mathrm{eq}l_B$,
and, adopting $l_B=l_0\Rmcr^{-1/2}$,
\begin{eqnarray}\label{sigmaRM}
\sigma_\RM&\simeq& K\nel B_\mathrm{eq}\Rmcr^{-1/2}(l_0L)^{1/2}\nonumber\\
&\simeq& 110\,{\displaystyle\frac{\rm{rad}}{\rm{m}^2}}
                \left(\frac{\nel}{10^{-3}\cmcube}\right)
                \left(\frac{B_\mathrm{eq}}{3\mkG}\right)\!
                \left(\frac{\Rmcr}{35}\right)^{-1/2}
\nonumber\\
        &&\mbox{}\times
                \left(\frac{l_0}{100\kpc}\right)^{1/2}
                \left(\frac{L}{750\kpc}\right)^{1/2}\!\!\!.
\end{eqnarray}
If the magnetic sheets multiply cover the projected area a turbulent cell,
by a factor
$q=O(1)$, then additional factor
$q^{1/2}$ has to be included in Eq.~(\ref{sigmaRM}).
The comparison
of Eq.~(\ref{sigmaRM}) with both numerical simulations and observations of
Faraday rotation in galaxy clusters suggest that
$q\approx1$.
If $B_\mathrm{eq}\propto v_0\propto t^{-3/5}$
and $l_0\propto t^{2/5}$, the observed $\RM$ will, on average, decrease with
time as
\begin{equation}\label{sRMt}
\sigma_\RM\propto [(t-t\f)/t_0\ini]^{-2/5}\;.
\end{equation}

\begin{figure}
\centerline{\includegraphics[width=0.495\textwidth]{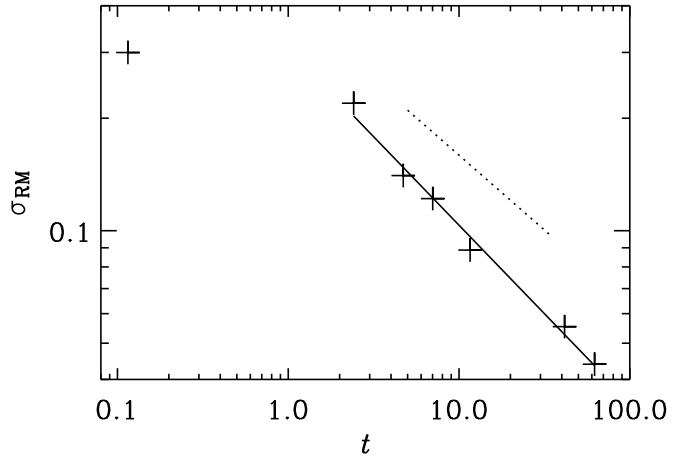}}
\caption{\label{RM_width}The width of the histogram of
Faraday rotation measures, $\sigma_{\rm RM}$, calculated along $256^2$ lines
of sight through the computational box, with $\tilde{k}\f=5$
(Model~1,  as in
Fig.~\ref{decay}), as a function of time. The solid line is the
least square fit to the data points at $t>1$, while the dotted line
corresponds to $t^{-2/5}$. Here $\sigma_\RM$ is measured in the units
$K\nel\mean{B^2}^{1/2} l_0\approx280\radm$, so that
$\sigma_\RM\approx80\radm$ in the steady state.
}
\end{figure}

We have calculated the Faraday rotation measure for
$256^2$ lines of sight through our computational domain, and then
computed $\sigma_\RM$ as the standard deviation of the results. The evolution
of $\sigma_\RM$ in decaying turbulence is shown
in Fig.~\ref{RM_width} for Model 1
and it exhibits remarkable agreement with Eq.~(\ref{sRMt}).
At earlier stages of the simulations in Model~1, $\sigma_\RM$
first grows rapidly while magnetic field is exponentially amplified, and then
remains fairly constant, $\sigma_\RM\approx0.3$, for $-20<t<1$
(Fig.~\ref{RM_width}). In dimensional
units, this corresponds to $\sigma_\RM\approx80\radm$. Model~2
results in a value of $\sigma_\RM\approx0.47$
 (Fig.~\ref{histogram2_Pm=30}), corresponding to $130\radm$.

\begin{figure}
\centerline{\includegraphics[width=0.495\textwidth]{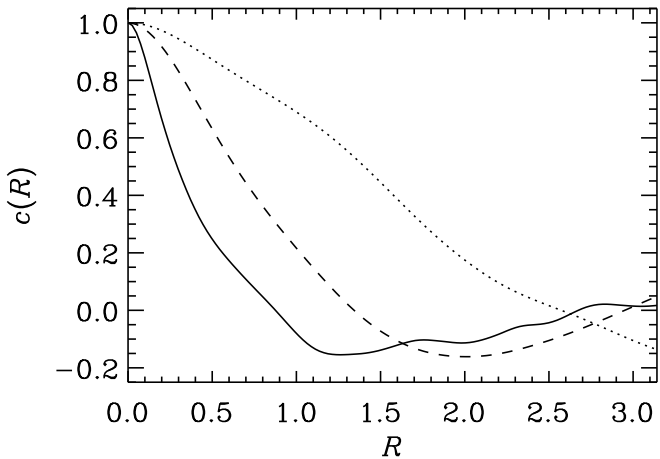}}
\caption{\label{structure_t}
The normalized autocorrelation function
$c(R)=C(R)/\sigma_\RM^2$, where
$C(R)=\langle\RM(\vect{X}+\vect{R})\RM(\vect{X})\rangle$ and
$\sigma_\RM^2=\langle \RM^2(\bmath{X})\rangle$ obtained for
Model 2 ($\tilde{k}\f=1.5$) with $\Pm=1$, at various times:
$t=0$ (solid), $t=30$ (dashed) and $t=70$ (dotted). The
former curve
refers to the statistically steady state, whereas the latter two
illustrate how $\RM$ distribution becomes less intermittent as turbulence
decays. Here $R$ is measured in the units of $\tilde{k}\f l_0/(2\pi)\approx70\kpc$.
}
\end{figure}

It is useful to compare results of the simulations with
the analytical estimate of Eq.~(\ref{sigmaRM}),
$\tilde{\sigma}_\RM \simeq (B_\mathrm{eq}/\mean{B^2}^{1/2})
\Rmcr^{-1/2} \tilde{k}\f^{1/2}$ in dimensionless units used in
Fig.~\ref{RM_width}.
This gives $\tilde{\sigma}_\RM \simeq 0.4$ for $\tilde k\f = 1.5$ in Model~2
and $\tilde{\sigma}_\RM \simeq 0.75$ for $\tilde k\f = 5$ in Model~1.
This estimate of $\tilde{\sigma}_\RM$ for Model~2,
which has a higher spatial resolution, is
in good agreement with the numerical simulations,
but that obtained for
Model~1 is a factor of about 2 lower
than expected.
Nevertheless,
our simulations confirm that
Eq.~(\ref{sigmaRM}) and Table~\ref{res} provide reasonably
good estimates of the expected amount
of Faraday rotation by magnetic field generated by the fluctuation dynamo.

Altogether the estimate (\ref{sigmaRM}) and the amount of Faraday rotation in
our simulations agree very well with observations of Faraday rotation in the
intracluster gas.

Figure~\ref{structure_t} shows the autocorrelation function of
the Faraday rotation measure for Model 2 at the beginning of the evolution
and at two later times.
As turbulence decays and magnetic field becomes less structured,
the correlation scale of $\RM$ fluctuations increases.
In the steady state, the dimensional value of the Taylor microscale
(or differential length scale) of the
$\RM$ fluctuations is, as expected, about
$R_\RM\approx15\kpc$, i.e., about half the
thickness of magnetic sheets, $l_B$, as quoted in Table~\ref{res}.

The situation is somewhat different for the wakes of subclusters and
individual galaxies. As discussed in Sects~\ref{SW} and \ref{GW}, their
volume filling factor is small, whereas the area covering factor can be of
order unity. In other words, a line of sight typically passes through just a single
turbulent wake, where the turbulent scale is comparable to the wake width. The
resulting Faraday rotation measure is given by
\begin{eqnarray}\label{sigmaRMw}
\sigma_\RM&=&K\nel B_\mathrm{eq} l_B\nonumber\\
&\simeq& 8.6\,\frac{\mbox{rad}}{\mbox{m}^2}
                \left(\frac{\nel}{10^{-3}\cmcube}\right)
                \left(\frac{B_\mathrm{eq}}{1\mkG}\right)
                \left(\frac{l_B}{10\kpc}\right),
\end{eqnarray}
where we retain the notation $\sigma_\RM$ because the resulting amount of
Faraday rotation will remain random, both because of the random orientation
of the wakes and due to the randomness of magnetic field within the wake.

Estimates of typical Faraday rotation measures obtained from Eqs~(\ref{sigmaRM})
and (\ref{sigmaRMw}) are given in the last column of Table~\ref{res}.

Schekochikhin et al.\ (2005a,b) suggest a different spatial structure of the
cluster magnetic field, based on their interpretation of the
fluctuation dynamo for $\Pm\gg1$
 (Schekochihin et al.\ 2004). These authors suggest that
magnetic field produced by the dynamo is locally anisotropic and represents
$l_0$-long magnetic sheets and/or ribbons multiply folded at a
microscopic resistive scale $l_\eta$. Such a model would typically
produce a much smaller $\sigma_\RM$.
Arguments similar to those that lead to
Eq.~(\ref{sigmaRM}),
(even assuming that the folds are randomised by some unspecified mechanism)
predict $\sigma_\RM$ smaller
than in our model by a factor
$F \simeq (\Rm/\Rmcr)^{1/2}$. In our simulations,
this corresponds to $F\simeq 3$--5 over the range $\Rm=400$--1000.
So the simulations that we have analysed (with $\Rm \simeq 400$),
cannot help to confidently discriminate between the
two magnetic field geometries.

However, observations indicate that magnetic coherence scale is at least
a few kpc and more plausibly exceed 10 kpc. This would be difficult to produce
in the model of Schekochihin et al.\ (2005a,b), unless the effective value of
$\Rm$ in the clusters is reduced, say due to plasma effects, to be
close to $\Rmcr$. Furthermore, Schekochihin et al.\ (2005a,b) envisage 
systematic reversals of the folded magnetic field along the line 
of sight, rather than random changes of its direction. Such a systematic 
behaviour would reduce $\sigma_\RM$ even further due to systematic 
cancellations of magnetic field along the line
of sight which would preclude the random walk of the polarization angle assumed
in Eq.~(\ref{sigmaRM}).

\begin{figure}
\centerline{\includegraphics[width=0.495\textwidth]{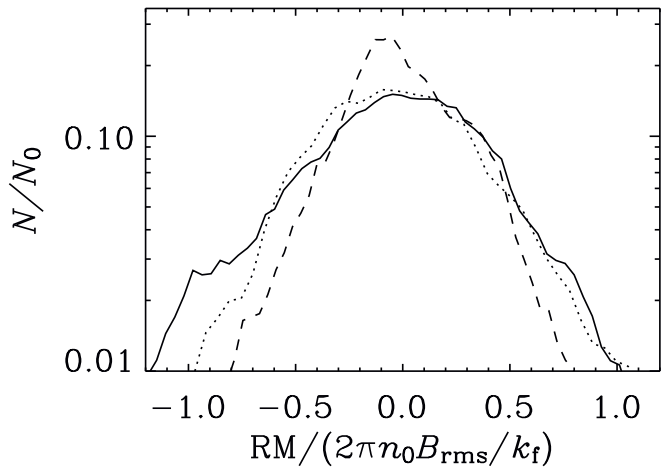}}
\caption{\label{histogram2_Pm=30}
The histogram (probability density) of the Faraday rotation measure calculated
along $256^2$ lines of sight through the computational box, with
 $\tilde{k}\f=1.5$, for $\Pm=1$ (solid),  $\Pm=30$ (dashed) and $\Pm=1/4$
(dash-dotted), all in the statistically steady state before the decay starts,
and with $\Rey= 450,44,445$ respectively.
Here $\RM$ is normalized by $K\nel\mean{B^2}^{1/2} l_0$, with
$B_\mathrm{rms}\equiv\mean{B^2}^{1/2}$.
}
\end{figure}

We have also used numerical simulations to examine the effects of varying the
magnetic Prandtl number on magnetic field structure and Faraday rotation.
The probability distribution of the Faraday rotation measure along $256^2$
lines of sight through the computational box is shown in
Fig.~\ref{histogram2_Pm=30} for three values of the magnetic Prandtl number.
The shape of the probability distribution is close to a Gaussian curve for
$\Pm=1/4$ and $\Pm=1$ (which is a parabolic shape in this representation),
but the distribution obtained at $\Pm=30$ exhibits
shorter tails at large $|\RM|$. The reason for
this is apparently the abundance of small-scale structures
that produce smaller Faraday rotation when $\Pm\geq1$,
i.e., when the magnetic dissipation scale is smaller. Nevertheless, the
standard deviation of the Faraday rotation measure has similar values for
both $\Pm=1$ and $\Pm=30$, $\sigma_\RM\approx0.47$ and $0.3$ in the units
of Fig.~\ref{histogram2_Pm=30}, respectively. This implies that magnetic field
does not become more strongly folded as $\Pm$ increases.
For the reader's convenience, we note again that $\Rey\approx44$ and
$\Rm\approx1300$ for $\Pm=30$ and  $\Rey\approx445$ and
$\Rm\approx111$ for $\Pm=1/4$.

\begin{figure}
\centerline{\includegraphics[width=0.495\textwidth]{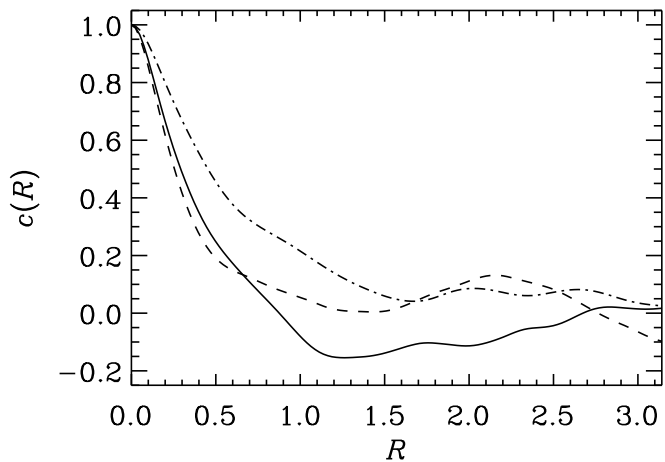}}
\caption{\label{structure}
As in Fig.~\ref{structure_t}, but for different magnetic Prandtl numbers;
$\Pm=1$ (solid), $\Pm=30$ (dashed) and
$\Pm=1/4$ (dotted), all in the statistically steady state
 with $\tilde{k}\f=1.5$ before the decay starts.
}
\end{figure}

Figure~\ref{structure} shows the autocorrelation function of the
Faraday rotation measure for
various values of $\Pm$.
The correlation scales for both $\Pm=30$ and $\Pm=1$ are comparable,
but that which obtains when $\Pm=1/4$ is
a factor of two larger.

We emphasize
again
 that both the form of the $\RM$ probability distribution and
the correlation function do not
change much as $\Pm$ increases from unity to 30. This suggests that our
results can be robust and directly comparable with observations even though
$\Pm\gg1$ in the intracluster gas.
Nevertheless, it
would be important to clarify this issue further using
simulations with higher resolution.
Note that the Faraday rotation measure
is proportional to the product of magnetic field and its scale; therefore, it
remains dominated by large scales
for magnetic spectra
$M_k\propto k^\kappa$ with $\kappa<1$.

\subsection{Polarization of cluster radio halos}
Estimates of the scale of magnetic field in galaxy clusters obtained above
are somewhat larger than what is usually adopted. We predict that the
number of turbulent correlation cells along a path length of $750\kpc$ can be only
about 3--5. As discussed above, magnetic
field in each cell has an intermittent component of randomly oriented
sheets and  somewhat weaker fields in a volume filling
component (see Fig.~\ref{snapshots}). It can be expected that synchrotron
emission produced in such a random
magnetic field will
be significantly polarized (assuming that magnetic field is well
ordered within individual
magnetic sheets). As discussed by Sokoloff et
al.\ (1998, their Sect.~5.1), the expected degree of polarization is a random
quantity whose standard deviation is about
\[
\sigma_p\simeq p_0 N^{-1/2}\;,
\]
where $p_0\approx70\%$ is the intrinsic degree of polarization and $N$ is the
number of turbulent cells within the beam cylinder, which yields $\sigma_p\simeq30\%$, neglecting
beam depolarization (see below). (We note that this estimate strictly applies
if $N\gg1$.) The polarized emission would be confined
to elongated structures
(cross-sections of magnetic sheets) of $l_B=20$--$40\kpc$ in width and
$l_0=150$--$300\kpc$ in length.
The intrinsic polarization plane should be perpendicular to the major axes of
the elongated synchrotron structures since magnetic field is mostly parallel to
the magnetic sheets.

However, the fractional polarization observed from cluster radio halos is less
than 2--10\% at the wavelength $\lambda=21\cm$  (L.~Feretti 2003, private
communication; Govoni \& Feretti 2004). No significant diffuse polarized emission
in the Coma cluster has been detected by Thierbach, Klein \& Wielebinski (2003)
at wavelengths $\lambda11.2\cm$ and $\lambda6.2\cm$. The depolarization can be attributed
to internal Faraday dispersion by the random magnetic field, where the
degree of polarization will be further reduced to [Eq.~(34) of Sokoloff et al.\ 1998]
\[
p=\sigma_p \frac{1-\exp(-S)}{S}\;,\qquad S=2\lambda^4\sigma_\RM^2
\]
(which is strictly applicable when $N\gg1$). Faraday dispersion readily
explains the lack of polarization at $\lambda21\cm$ where this equation yields
$p\approx0.2\%$ for $\sigma_\RM=200\radm$. The Faraday
depolarization is weaker at shorter wavelengths, with $p\approx3\%$ at
$\lambda11\cm$ and $p\approx20\%$ at $\lambda=6\cm$. However, the linear
resolution of the observations of Thierbach et al.\ (2003) was $W=110\kpc$ at
$\lambda11.2\cm$ and $W=60\kpc$ at $\lambda6.2\cm$. Given that the thickness
of the elongated polarized structures is of order $l_B=25\kpc$, beam
depolarization would further reduce the degree of polarization at least by a
factor $W/l_B$ to $0.5\%$ at $\lambda11.2\cm$ and 8\% at $\lambda6\cm$. These
estimates indicate that the
polarization of cluster synchrotron halos should  be
weak but detectable at sufficiently high resolution
and short wavelengths. In reality, each
correlation cell may contain a few magnetic sheets with independent directions
of magnetic field (cf.\ Fig.~\ref{snapshots}). Therefore, the effective number
of magnetic sheets along the path length
(and/or within the telescope beam)
 can be a factor 2--3 larger than
adopted above and our values of the degree of polarization
can be overestimated by a factor of two. Further polarization
observations of cluster radio halos at short wavelengths can reveal magnetic
structures suggested here.

Shear in the gas motions at a scale of a few hundred kiloparsecs, produced
during major merger events, can make the random magnetic field locally
anisotropic. The anisotropy can also be produced by differential rotation
and/or inhomogeneous inflow in cluster cores. Anisotropic random magnetic
field can produce significant polarization of the synchrotron emission, with
the polarization vector orthogonal to the direction of the maximum r.m.s.\
field strength (Laing 1981; Sokoloff et al.\ 1998). This polarization
can be observable if the shear regions are large enough as to avoid the
cancellation of polarization along the line of sight
or across the telescope beam.

Govoni et al.\ (2005)
report detection of polarized emission from filamentary structures in the
cluster A2255, of a size $200\kpc\times500\kpc$ (see also Murgia et al.\
2004), but the orientation of the polarization plane mostly disagrees with
the above suggestions,
 unless the amount of foreground Faraday rotation is
larger than that assumed by Govoni et al.

Another situation where significant polarization of synchrotron emission in the
cluster environment can be expected are the wakes of subclusters and galaxies.
Since  individual lines of sight pass through one (or a few) wakes wherein the
turbulent scale is comparable to the wake width,
polarization due to the
random magnetic field can be detectable.

\section{Discussion}\label{Disc}
There is growing direct and indirect evidence for the presence of random --
and plausibly turbulent -- motions in the intergalactic gas of galaxy
clusters. We have identified several stages in their evolution, from a
statistically quasi-steady motion during the epoch of major mergers, to the
stage of decaying turbulence that follows, and to a state where
turbulence is confined to the wakes of relatively small subclusters and
individual galaxies. Typical parameters of the velocity and magnetic fields
at various stages of the cluster evolution are summarized in Table~\ref{res}:
random velocities of $v_0=150$--$300\kms$  can be maintained at various stages
of the evolution, and their scale is expected to be $l_0=150$--$300\kpc$, with
the exception of galactic wakes where it can be of
the order of $10\kpc$.

It is not quite clear whether or not the random motions in the intracluster
gas can evolve into developed turbulence. This depends on the value of the
Reynolds number, a measure of the relative strength of nonlinear
hydrodynamic effects and, therefore, of the strength
of the spectral energy cascade. If the flow remains
laminar, the motions can decay faster after the end of the major mergers,
and the wakes can have properties different from those discussed above.
However, our numerical simulations suggest that
the power-law decay establishes itself
even for the Reynolds number as modest as $\Rey\simeq100$.

The turbulent flow of magnetized gas
 can accelerate relativistic particles required to produce
cluster radio halos (Tribble 1993b; Brunetti et al.\ 2004; Cassano \& Brunetti
2005). Turbulent mixing in clusters has also been invoked in
modeling the  transport of heat
(Cho et al.\ 2003; Kim \& Narayan 2003; Voigt \& Fabian 2004;
Dennis \& Chandran 2005), and metals (Rebusco et al.\ 2005). The dissipation
of the turbulent energy can help
to balance the cooling of cluster cores
(Fujita, Matsumoto \& Wada 2004; Rebusco et al.\ 2005).
Clearly, turbulence in clusters seems to be useful to understand
several diverse aspects of cluster physics.

Random motions during and immediately after the major merger epoch are
plausibly volume-filling. However, random flows produced by the wakes can have
area covering factor of order unity, but the volume filling factor of such
wakes can be relatively small, $f_V\simeq0.1$, leaving large quiescent regions
between the wakes. Therefore, a typical line of sight passes through a
turbulent region with an r.m.s.\ speed of a 200--$300\kms$, producing
some observational signatures of developed turbulence, and yet there is enough
space to accommodate well-ordered morphological features apparently
unaffected by any random motions.
For example, in the core of the Perseus
cluster, where the density is higher (Churazov et al.\ 2004),
with $\nel \geq 10^{-2}$ cm$^{-3}$ for $r \leq 100$ kpc, the mean free path is
smaller $\lambda \leq 0.5$ kpc, and the wakes can have a potentially
larger filling factors $f_V \leq 0.5$, for $\delta =0.2$.
Thus the presence of long H$\alpha$
filaments observed by Fabian et al.\ (2003, 2005) in the core of the Perseus
cluster may not be inconsistent with various evidence for random motions
in this cluster core (Churazov et al.\ 2004; Rebusco et al.\ 2005).
A possible signature of such spatially intermittent turbulence  could be a
specific shape of spectral lines, with a narrow core, produced in quiescent
regions, accompanied by nonthermally broadened wings.
It would be interesting to pursue this idea further in a more
quantitative fashion.

In the presence of random motions, any pre-existing magnetic field will be
rapidly destroyed owing to a reduction of its scale by the velocity shear.
Even in a quiescent medium, any nonuniform magnetic would decay by driving
motions whose kinetic energy can be efficiently converted into heat because
the intracluster gas is expected to be rather viscous. Therefore, random
magnetic fields confidently revealed in many clusters through their Faraday
rotation must be constantly maintained even if the electrical conductivity of
the intracluster gas is large.

However, the same random motions -- either turbulent or not -- will generate
magnetic fields via the fluctuation dynamo action at all the stages of the
cluster evolution. The field is amplified by random shear, which reduces
its scale along the directions perpendicular to
 the shear layers.  This makes the spatial distribution of
the magnetic field intermittent.
Numerical simulations give an impression of strong field
regions being largely confined into magnetic sheets and ribbons
(and, with lower probability, filaments)
wherein its strength is similar to that given by energy equipartition
with the overall kinetic
energy density. Following Subramanian (1999), we argue that the volume filling
factor of the magnetic structures within a turbulent cell (provided they are
mostly sheets rather than filaments) is of order 0.1--0.2. Our numerical
simulations confirm this picture, but add to it a weaker volume-filling
magnetic background, so that the total magnetic energy density is about
1/5--1/3 of the kinetic energy density of the random flow.
The limited experience available with fluctuation dynamo models at
large magnetic
Prandtl number seems to indicate that magnetic fields can be closer to
energy equipartition with turbulence in more realistic models.

The (random) Faraday rotation
measures produced by such magnetic fields are in the $1\sigma$ range of
100--$200\radm$ in agreement with observations. We note, however, that,
according to our estimates, the scale of the magnetic field is
$l_B=20$--40\,kpc, i.e., a factor of a few larger than what is usually
assumed.
The scale of the field is smaller, of order
a $\kpc$, for galactic wakes;
it could also be smaller if there were other sources of stirring like
radio galaxies.
The maximum field strength in the magnetic structures is about
2--$4\mkG$, whereas its r.m.s.\ value within a turbulent cell is 1--$2\mkG$.
Such r.m.s.\ field strengths are in better accord with those inferred from
synchrotron intensity assuming equipartition between
magnetic fields and cosmic rays, or with inverse Compton limits.

We predict that synchrotron emission from cluster radio halos similar to that
in the Coma cluster can be significantly polarized at short wavelengths
$\lambda=3$--5\,cm.

\section*{Acknowledgments}
We are grateful to R.~Beck, A.~Brandenburg, A.~C.~Fabian, L.~Feretti and
M.~J.~Rees for useful discussions. We appreciate detailed comments
of the referee, A.~Schekochihin, which helped to improve the
presentation. This work
was supported by IUCAA, a joint grant program of the Indian National Science
Academy and the Royal Society, and PPARC Grant PPA/G/S/2000/00528. AS and KS
acknowledge the hospitality and partial financial support of the Isaac Newton
Institute for Mathematical Sciences, Cambridge. AS is grateful to
Carlo Barenghi, Andrew Fletcher and Graeme Sarson
who took over his teaching responsibilities during his stay in Cambridge.


\appendix

\section{Seed magnetic fields in
galaxy clusters}\label{seed}
There is a number of sources of seed magnetic fields
in galaxy clusters. It is well known that the intracluster medium (ICM)
has
high metallicity which must have been produced in stars in galaxies
and subsequently ejected into the galactic interstellar medium (ISM)
and then into the ICM. Since the ISM is likely to be magnetized with
fields of order a few $\mu$G, this would lead to a seed
field in the ICM. One can
roughly estimate the seed field resulting from stripping the galactic gas, by
using magnetic flux conservation under spherically symmetric expansion;
that is $B_{\rm seed}\simeq(\rho_\mathrm{ICM}/\rho_\mathrm{ISM})^{2/3}
B_\mathrm{gal}$. For $ B_\mathrm{gal} \simeq 3 \mkG$, and
$\rho_\mathrm{ICM}/\rho_\mathrm{ISM} \simeq 10^{-2}$--$10^{-3}$, one gets
$B_\mathrm{seed} \simeq0.1$--$0.03\mkG$. One may get even larger seed fields if
there is a
substantial number of active galaxies with magnetized outflows: if
about $10^3$
galaxies have mass outflow with
$\dot{M} \simeq 0.1M_\odot \yr^{-1}$ lasting for
$1\Gyr$, with a Poynting flux about 10\% of the material flux, and the field
gets mixed into the cluster gas over a Mpc sized region, $B_\mathrm{seed} \simeq
0.3\mkG$ would result (Brandenburg 2000). This estimate, however, assumes
that all the intracluster gas has been processed through the outflows,
which may be an exaggeration.

Another source of seed fields is likely to be the outflows from earlier
generation of active galaxies (radio galaxies and quasars) (Rees 1994;
Goldshmidt \& Rephaeli 1994; Medina-Tanco \& En{\ss}lin 2001; Furlanetto \&
Loeb 2001; Colgate, Li \& Pariev 2001). Such outflows can produce magnetized
plasma bubbles in some fraction of the intergalactic volume (typically of order
$10\%$ -- Furlanetto \& Loeb 2001) which, when incorporated into the ICM,
would seed the general cluster gas with magnetic fields. If one assumes that
the cluster gas is $10^3$ times denser than the
intergalactic medium and blindly uses the
enhancement of the bubble field due to compressions during cluster formation,
one can get fields as large as $0.1$--$1\mkG$ in the ICM (Furlanetto \& Loeb
2001). However this is to ignore the issue of how the field in the magnetized
bubble, especially if it is predominantly relativistic plasma from a radio
galaxy, mixes with the unmagnetized and predominantly thermal gas during
cluster formation, and the resulting effects on both the field strength and
coherence scale
 (see En{\ss}lin 2003 for the related problem of the escape of
cosmic rays out of radio cocoons). It is likely that, while AGNs and galaxies
provide a potentially strong seed magnetic field, there would still be a need
for their subsequent amplification and maintenance against turbulent decay.

Altogether, we adopt $B_\mathrm{seed}=10^{-7}\G$ as a plausible estimate of the
seed magnetic field in the intracluster gas.

\section{The correlation function of
the
Faraday rotation measure}\label{sRM}
In order to calculate the autocorrelation function of the Faraday rotation
measure, defined in Eq.~(\ref{RMdef}), we introduce  coordinates $(x,y,z)$
with the $z$-axis directed towards the observer, and those in the plane of the
sky, $\vect{X}=(X,Y)$. We assume the magnetic field to be an isotropic,
homogeneous, random field with zero mean value. Then its equal-time,
two-point correlation tensor has the form $\mean{B_i(\vect{x},t)
B_j(\vect{y},t)} = M_{ij}(r,t)$, where
\[
M_{ij} =
\left(\delta_{ij} -\frac{r_i r_j}{r^2}\right) M_\mathrm{N}(r,t)
+\frac{r_i r_j}{r^2} M_\mathrm{L}(r,t)\;.
\]
Here $\mean{\cdots}$ denotes the ensemble average, $r=|\vect{x}-\vect{y}|$,
and $r_i = x_i -y_i$ (see Section~34 of Landau \& Lifshitz
1975; Monin \& Yaglom 1975). The functions $M_\mathrm{L}(r,t)$ and
$M_\mathrm{N}(r,t)$ are known as the longitudinal and transverse correlation
function of the magnetic field, respectively. Since $\nabla\cdot\vect{B}=0$,
\[
M_\mathrm{N} = {1\over2r}{\partial\over\partial r}\left(r^2 M_\mathrm{L}\right).
\]
We also assume for simplicity that the electron density
is uncorrelated with the magnetic field
and also constant over the field correlation length.
This is consistent with the fact that random gas motions in galaxy
clusters are quite subsonic. The correlation function of $\RM$ is then
\begin{eqnarray}
C(R) &=& \mean{\RM(\vect{X}_1)\RM(\vect{X}_2)}\nonumber\\
&=&K^2\nel^2\int_0^L \int_0^L \mean{B_z(\vect{X}_1,z_1)B_z(\vect{X}_2,z_2)}\,dz_1\,dz_2\nonumber\\
&=&K^2\nel^2 L \int_{-L}^L M_{zz}(R,\zeta) d\zeta \nonumber \\
&=& K^2\nel^2 L \int_{-L}^L
\left(M_\mathrm{L} \frac{R^2}{R^2+\zeta^2} +
M_\mathrm{N} \frac{\zeta^2}{R^2+\zeta^2} \right)d\zeta\nonumber\\
&=& K^2\nel^2 L \int_{-L}^L
\left(M_\mathrm{L} +\frac{\zeta^2}{2r}\frac{dM_\mathrm{L}}{dr}\right)d\zeta\;.
\label{cR}
\end{eqnarray}
Here we have assumed that $L$ is much greater than the correlation length of
the magnetic field, $\zeta=z_1-z_2$, $R=|\vect{X}_1 - \vect{X}_2|$ and
$r^2=R^2+\zeta^2$.

For the sake of illustration, consider the longitudinal correlation function of
the form
\[
M_\mathrm{L} = \sfrac13\mean{B^2}\exp \left( -\frac{r^2}{2l_B^2} \right),
\]
which corresponds to the one-dimensional magnetic spectrum of the form
$M_k\propto k^4\exp{(-k^2l_B^2/2)}$ (Monin \& Yaglom 1975). We note that
$M_k$ attains maximum at
a wavenumber $k_m = 2/l_B$, (or a scale
$2\pi/k_m = \pi l_B$), whereas the longitudinal
correlation scale is given by
$l_\mathrm{L}=[M_\mathrm{L}(0)]^{-1}\int_0^\infty M_\mathrm{L}(r)\,dr=l_B\sqrt{\pi/2}$.

Straightforward calculation then yields
\begin{equation}\label{CRG}
C(R)=\frac{\sqrt2}{3}c
K^2\nel^2
\mean{B^2} Ll_B\exp\left( -\frac{R^2}{2l_B^2} \right),
\end{equation}
where
\[
c = \int_{-L/(\sqrt{2}l_B)}^{L/(\sqrt{2}l_B)} (1 - s^2)\exp{(-s^2/2)}\,ds
\approx 0.88\;,
\]
with the numerical value obtained for $L/(\sqrt{2}l_B) \gg 1$.

The r.m.s. value of $\RM$ can be obtained from Eq.~(\ref{cR}) or (\ref{CRG}) at
$R=0$:
\begin{eqnarray}
\sigma_\RM^2 &=& K^2\nel^2 L \int_{-L}^L\left. M_\mathrm{N}(R,z)\right|_{R=0}\, d\zeta\nonumber\\
&=&\frac{\sqrt2}{3}c K^2\nel^2 \mean{B^2} Ll_B\;,
\label{c0}
\end{eqnarray}
which is similar to Eqs~(\ref{sRMN}) and (\ref{sigmaRM}).

\label{lastpage}
\end{document}